\documentclass[twocolumn,showpacs,aps,prb]{revtex4}
\usepackage{epsfig}
\usepackage{hyperref}
\usepackage{amsmath,amssymb}
\usepackage{graphicx}
\usepackage{dcolumn}
\usepackage{graphics}

\begin{document}
\title{AC-driven vortices and the Hall effect in a tilted washboard planar pinning potential}

\author{Valerij~A.~Shklovskij}
\address
{Institute of Theoretical Physics, National Science Center-Kharkov
Institute of Physics and Technology, 61108, Kharkov,
Ukraine;\\Kharkov National University, Physical Department, 61077,
Kharkov, Ukraine}
\author{Oleksandr~V.~Dobrovolskiy}
\address {Kharkov National University, Physical Department, 61077, Kharkov,
Ukraine}\date{\today}


\begin{abstract}
The Langevin equation for a two-dimensional (2D) nonlinear guided
vortex motion in a tilted cosine pinning potential in the presence
of an \emph{ac} current is \textit{exactly} solved in terms of a
\textit{matrix} continued fraction at arbitrary value of the Hall
effect. The influence of an \emph{ac} current of arbitrary
amplitude and frequency on the \emph{dc} and \emph{ac}
magnetoresistivity tensors is analyzed. The  $ac$ current density
and frequency dependence of the overall shape and the number and
position of the Shapiro steps on the anisotropic current-voltage
characteristics is considered. An influence of a subcritical or
overcritical \emph{dc} current on the time-dependent stationary
\emph{ac} longitudinal and transverse resistive vortex response
(on the frequency of an \emph{ac}-driving $\Omega$) in terms of
the nonlinear impedance tensor $\hat{Z}$  and a nonlinear
\emph{ac} response at $\Omega$-harmonics are studied. New
analytical formulas for $2D$ temperature-dependent \textit{linear}
impedance tensor $\hat{Z}_L$ in the presence of a $\emph{dc}$
current which depend on the angle $\alpha$ between the current
density vector and the guiding direction of the washboard PPP are
derived and analyzed. Influence of $\alpha$-anisotropy and the
Hall effect on the nonlinear power absorption by vortices is
discussed.
\end{abstract}
 \maketitle


\section{Introduction}

It is well known that the mixed-state resistive properties of
type-II superconductors are determined by the dynamics of vortices
which in the presence of pinning sites may be described as a
motion of vortices in some pinning potential~\cite{BLAT}. In the
simplest case this pinning potential is assumed to be periodic in
one dimension, and temperature-dependent $dc$ current uniaxial
pinning anisotropy, provoked by such washboard planar pinning
potential (PPP) recently has been extensively studied both
theoretically$^{2-4}$ and experimentally.$^{5-9}$ Two main reasons
stimulated these studies. First, in some high-$T_c$
superconductors (HTSCs) twins can easily be formed during  the
crystal growth.$^{5-7}$ Second, in layered HTSCs the system of
interlayers between parallel $ab$-planes can be considered as a
set of unidirectional planar defects which provoke the intrinsic
pinning of vortices.~\cite{BLAT}

As the pinning force in a PPP is directed perpendicular to the
washboard channels of the PPP~\cite{BLAT}, the vortices generally
tend to move along these channels. Such a \emph{guided} motion of
vortices in the presence of the Hall effect produces anisotropic
transport behaviour for which even (+) and odd (--) (with respect
to the magnetic field reversal) longitudinal ($\parallel$) and
transverse ($\perp$) $dc$ nonlinear magnetoresistivities
$\rho_{\parallel,\perp}^{dc\pm}$ depend substantially on the angle
$\alpha$ between the $dc$ current density vector $\mathbf{j}$ and
the direction of the PPP channels ("guiding direction").

The $dc$-current nonlinear guiding problem was exactly solved
recently for the washboard PPP within the framework of the
two-dimensional ($2D$) single-vortex stochastic model of
anisotropic pinning based on the Fokker-Planck equation and rather
simple formulas were derived for the $dc$ magnetoresistivities
$\rho_{\parallel,\perp}^{\pm}$.\cite{MAW,SSS}

On the other hand, the high-frequency and microwave
\textit{impedance} measurements of a mixed state can also give
information about the flux pinning mechanisms and the vortex
dynamics. One of the most popular experimental methods for the
investigation of the vortex dynamics in type-II superconductors is
the measurement of the complex $ac$ response in the radiofrequency
and microwave ranges. When the Lorentz force acting on the
vortices is alternating, then due to the pinning the $ac$
resistive response acquires imaginary (out-of-phase) component.
Due to this reason measurements of the complex $ac$ response
versus frequency $\omega$ can give important information on the
pinning forces.

The very early model of Gittleman and Rosenblum~\cite{GR} (GR)
considered oscillations of damped vortex in a garmonic pinning
potential. GR measured the power absorption of the vortices in
PbIn and NbTa films over a wide range of frequencies $\omega$ and
successfully analyzed their data with the simple equation
\begin{equation}
    \label{I1}
    \eta\dot{x} + k_px = F_L,
\end{equation}
where $x$ is the vortex displacement, $\eta$ is the vortex
viscosity, $k_p$ is the pinning constant, and $F_L$ is the Lorentz
force. From Eq.~\eqref{I1} follows that the complex vortex
resistivity $\rho_v$ is
\begin{equation}
    \label{I}
    (\rho_v/\rho_f) = i(\omega/\omega_p)/[1 + i(\omega/\omega_p)],
\end{equation}
where $\rho_f$ is the flux-flow resistivity and $\omega_p \equiv
k_p/\eta$ is the depinning frequency. As follows from
Eqs.~\eqref{I1} and~\eqref{I}, pinning forces dominate at low
frequencies ($\omega \ll\omega_p$) where $\rho_v$ is
nondissipative, whereas at high frequencies ($\omega \gg
\omega_p$) frictional forces dominate and the vortex resistivity
is dissipative.

The experimental success of this very simple model stimulated the
attempts to use it for the interpretation of the data taken in
HTCSs, where the effects of thermal agitation are especially
important due to their low pinning activation energies and the
high temperatures of the superconducting state. As the GR model
was developed for zero temperature and could not account for the
thermally activated flux flow and creep, which are very pronounced
in HTCSs, there was a need for a more general model for the $ac$
vortex dissipation at different temperatures and frequencies.

In order to fulfill this aim the vortex equation of
motion~\eqref{I1} was supplemented with Langevin force which was
assumed to be Gaussian white noise with zero mean and the cosine
periodic pinning potential was used$^{11-13}$ for taking into
account the possibility of vortex hopping between different
potential wells. In the limit of small $ac$ current (i.~e. for a
\textit{nontilted} cosine pinning potential) this new equation of
motion was approximately solved by a continued-fraction
expansion$^{11-13}$ using the analogy between a pinned vortex and
a Brownian particle motion in a periodic potential. As a result,
the complex resistivity $\rho_v$ which generalizes the GR`s
Eq.~\eqref{I} has the form (see Eq.~(8) in Ref.~11)
\begin{equation}
    \label{2}
    \quad(\rho_v/\rho_f) =[i(\omega/\omega_0) + \nu_{00}]/[1 +
    i(\omega/\omega_0)],
\end{equation}
where $\nu_{00}$ is a creep factor that grows monotonically with
temperature increasing from $\nu_{00}=0$ (no flux creep) to
$\nu_{00}=1$ (flux flow regime) and $\omega_0$ is a characteristic
frequency (nonmonotonic in temperature) which, in absence of
creep, corresponds to the depinning frequency $\omega_p$. If the
frequency $\omega$ is swept across the temperature-dependent
frequency $\omega_0$, the observed $\textrm{Re}\rho_v$ increases
from a low frequency value to the flux-flow value $\rho_f$ while
$\textrm{Im}\rho_v$ exhibits a maximum at $\omega_0$. Thus, we can
summarize that the temperature-dependent $ac$-driven vortex motion
problem has been \textit{exactly} solved so far only for the
one-dimensional (1D) \textit{nontilted} cosine pinning potential
at a small oscillation amplitude of the vortices.

At the same time, the examination of a \textit{strong}
$ac$-driving (that is interesting both for theory and for
different high-frequency or microwave applications) evidently
requires to consider strongly \textit{tilted} pinning potential.
Actually, if at low temperatures and relatively high frequencies
in \textit{nontilted} pinning potential each pinned vortex will be
confined to its pinning potential well during the $ac$ period, in
the case of strong $ac+dc$ driving current the \textit{running}
states of the vortex may appear when it can visit several (or
many) potential wells during the $ac$ period.

The aim of this work is to suggest a new theoretical approach to
the study of temperature-dependent \textit{nonlinear}
\textit{ac}-driven pinning-mediated vortex dynamics based on an
\textit{exact} solution (in terms of a \textit{matrix} continued
fraction) of the same equation of vortex motion, as was discussed
by Coffey and Clem (CC) in the seminal paper~\cite{CLEM} (see
below Eq.~\eqref{F1} which has an additional Hall term). This new
approach substantially generalizes the CC`s results because the
two-dimensional (2D) Langevin equation for the nonlinear guided
motion in a \textit{tilted} cosine PPP in the presence of a strong
$ac$ current at arbitrary value of the Hall effect has been
exactly solved. For this exact solution we used the matrix
continued fraction technique earlier suggested and later
extensively employed for calculation of $1D$ nonlinear
($ac+dc$)-driven response of overdamped Josephson junction with
noise in Refs.~14,~15.

As a result, two groups of new findings were obtained. First, for
previously solved in Refs.~2 and~3 the $2D$ $dc$-problem of the
influence of an $ac$ current on the overall shape and appearance
of the Shapiro steps on the anisotropic $dc$
$\rho_{\parallel,\perp}^{dc\pm}$ -- CVC`s was calculated and
analyzed. Second, for the $ac$ current at a frequency $\omega$
plus $dc$ bias the $2D$ nonlinear time-dependent stationary
$\rho_{\parallel,\perp}^{ac\pm}$ $ac$-response on the frequency
$\omega$ in terms of nonlinear impedance tensor $\hat{Z}$ and a
nonlinear \emph{ac} response at $\omega$-harmonics was studied.

The organization of the paper is as follows. In Sec.~II we
introduce the model and the basic quantities of interest, namely,
the average  two-dimensional electric field and the Fourier
amplitudes for the averaged moments $\langle r^m \rangle$. In
Sec.~III we present the solution of the recurrence equations for
the Fourier amplitudes in terms of matrix continued fraction and
introduce the main anisotropic nonlinear component of our theory
-- the average pinning force, divided into three parts. In Sec. IV
we discuss the $\omega$-dependent $dc$ current magnetoresistivity
response with different (from A to E subsections) aspects of this
problem. Section~V (with subsections from A to H ) represents
different problems related to nonlinear anisotropic stationary
\emph{ac} response. In Sec.~VI we conclude with a general
discussion of our results.


\section{Formulation of the problem}

The Langevin equation for a vortex moving with velocity
$\mathbf{v}$ in a magnetic field $\mathbf{B}=\mathbf{n}B$
($B\equiv|\mathbf{B}|$, $\mathbf{n}=n\mathbf{z}$, $\mathbf{z}$~is
the unit vector in the $z$ direction and $n=\pm 1$) has the form
\begin{equation}
    \label{F1}
    \eta\mathbf{v}+n\alpha_{H}\mathbf{v}\times\mathbf{z}=\mathbf{F}_{L}+\mathbf{F}_{p}+\mathbf{F}_{th},
\end{equation}
where $\mathbf{F}_{L}=n(\Phi_{0}/c)\mathbf{j}\times\mathbf{z}$ is
the Lorentz force ($\Phi_{0}$ is the magnetic flux quantum, and
$c$ is the speed of light),
$\mathbf{j}=\mathbf{j}(t)=\mathbf{j}^{dc}+\mathbf{j}^{ac}\cos
\omega t$, where $\mathbf{j}^{dc}$ and $\mathbf{j}^{ac}$ are the
$dc$ and $ac$ current density amplitudes and $\omega$ is the
angular frequency, $\mathbf{F}_{p}=-\nabla U_{p}(x)$ is the
anisotropic pinning force ($U_{p}(x)$ is the washboard planar
pinning potential), $\mathbf{F}_{th}$ is the thermal fluctuation
force, $\eta$ is the vortex viscosity, and $\alpha_{H}$ is the
Hall constant. We assume that the fluctuational force
$\mathbf{F}_{th}(t)$ is represented by a Gaussian white noise,
whose stochastic properties are assigned by the relations
\begin {equation}
    \label{F2}
    \langle F_{th,i}(t)\rangle=0, \, \langle F_{th,i}(t)F_{th,j}(t') \rangle=2T\eta\delta_{ij}\delta(t-t'),
\end{equation}
where $T$ is the temperature in energy units, $\langle...\rangle$
means the statistical average, and $F_{th,i}(t)$ with $i=x$ or
$i=y$ is the $i$ component of $\mathbf{F}_{th}(t)$.

\begin{figure}[t]
    \epsfig{file=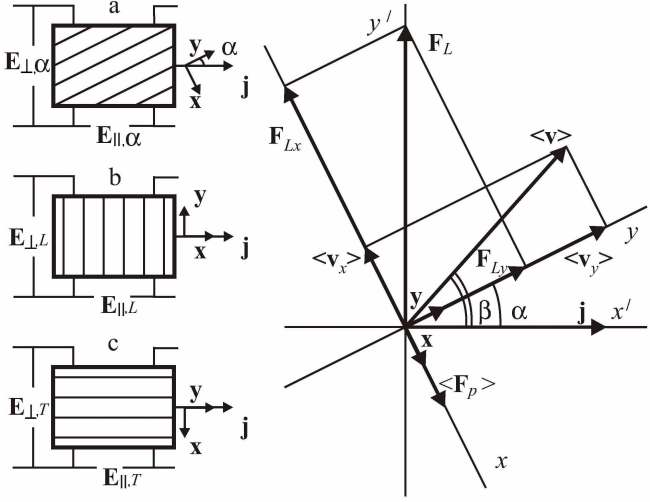,width=0.48\textwidth}
    \caption{System of coordinates $xy$ (with the unit vectors
$\mathbf{x}$ and $\mathbf{y}$) associated with the PPP washboard
channels and the system of coordinates $x'y'$ associated with the
direction of the current density vector $\mathbf{j}$; $\alpha$ is
the angle between the channels of the PPP and $\mathbf{j}$,
$\beta$ is the angle between the average velocity vector of the
vortices $\langle \textbf{v} \rangle$ and $\mathbf{j}$;
$\mathbf{F}_{L}$ is the Lorentz force;  $<\mathbf{F}_{p}>$ is the
average pinning force, $\mathbf{F}_{Lx}$ is the average effective
motive force for a vortex. Here for simplicity we assume
$\epsilon=0$. The schematic sample configuration for three cases
with different values of angle $\alpha$ (the insert): general
case, $\alpha\neq0, \pi/2$~(a); longitudinal $L$-geometry,
$\alpha=\pi/2$, $\mathbf{j}\perp \mathbf{y}$~(b); transverse
$T$-geometry, $\alpha=0$, $\mathbf{j}\perp \mathbf{x}$~(c); in all
cases $E_{\perp}$ and $E_{\parallel}$  are transverse and
longitudinal (with respect to $\mathbf{j}$-direction) electric
field components.} \label{fig1}
\end{figure}

The formal statistical average of Eq.~\eqref{F1} is
\begin {equation}
    \label{F3}
    \eta\langle\mathbf{v}\rangle+n\alpha_{H}\langle\mathbf{v}\rangle\times
    \mathbf{z}=\mathbf{F}_{L}+\langle\mathbf{F}_{p}\rangle.
\end{equation}
Though $\langle \mathbf{F}_{th}\rangle$ disappears because of the
stochastic property in Eq.~\eqref{F2}, effects of the thermal
fluctuation are implicit in the term $\langle \mathbf{F}_p
\rangle$ (see below).

Since the anisotropic pinning potential is assumed to depend only
on the $x$ coordinate and is assumed to be periodic
($U_p(x)=U_p(x+a)$, where $a$ is the period), the pinning force is
always directed along the anisotropy axis $x$ (with unity vector
$\textbf{x}$, see Fig.~1) so that it has no component along the
$y$ axis ($F_{py}=-dU_p/dy=0$). Thus, Eq.~\eqref{F1} reduces to
the equations
\begin{equation}
    \label{F4}
    \left\{
        \begin{array}{lll}
        v_x+\delta v_y=(F_{Lx}+F_{px}+F_x/)\eta,\\
        \\
        v_y-\delta v_x=(F_{Ly}+F_{y})/\eta,\\
        \end{array}
    \right.
\end{equation}
where $\delta\equiv n\epsilon$, $\epsilon\equiv \alpha_H/\eta$,
and we omitted index $th$ in the $\mathbf{F}_{th}$ for simplicity.
Eqs.~\eqref{F4} can be rewritten for the subsequent analysis in
the following form
\begin{equation}
    \label{F5}
    \begin{array}{crr}
    v_x\equiv \dot x=(\tilde{F}_{Lx}+F_{px}+\tilde{F}_x)/\eta D,\\
    \\
    v_y\equiv \dot y=(\tilde{F}_{Ly}+\tilde{F}_{y}+\delta F_{px})/\eta D.\\
    \end{array}
\end{equation}
where $\tilde{F}_{Lx}\equiv F_{Lx}-\delta F_{Ly}$,
$\tilde{F}_{Ly}\equiv F_{Ly}+\delta F_{Lx}$, $\tilde{F}_{x}\equiv
F_{x}-\delta F_{y}$, $\tilde{F}_{y}\equiv F_{y}+\delta F_{x}$,
$D\equiv1+\delta^2$ and $\langle \tilde{F}_i(t)
\tilde{F}_j(t')\rangle=2T\eta D\delta_{ij}\delta(t-t')$.

Our aim now is to obtain from Eqs.~\eqref{F5} a rigorous and
explicit expression of $\langle v_x \rangle$ and $\langle v_y
\rangle$ in which effects of the pinning and the thermal
fluctuation are considered. We assume, as usual$^{2, 11-13}$, a
periodic pinning potential of the form $U_p(x)=(U_p/2)(1-\cos
kx))$ where $k=2\pi/a$. As $F_{px}=-F_p \sin kx$, where $F_p\equiv
U_pk/2$, the first from Eqs.~\eqref{F5} has the form
\begin{equation}
    \label{F6}
    \hat{\tau}(d\textsl{x}/dt)+\sin\textsl{x}=\hat{F}_{Lx}+\hat{F}_x.
\end{equation}
Here $\textsl{x}=kx$ is the dimensionless vortex coordinate,
$\hat{\tau}\equiv\eta D/kF_p$ is the relaxation time,
$\hat{F}_{Lx}=\tilde{F}_{Lx}/F_p$ is the dimensionless generalized
moving force in the $x$ direction,
$\hat{F}_{x}=\tilde{F}_{x}/F_p$, and $\langle
\hat{F}_{x}(t)\hat{F}_{x}(t')\rangle=\tau\delta(t-t')$, where
$\tau\equiv\hat{2\tau}/g$ and $g=U_p/2T$ is the dimensionless
inverse temperature.

Making the transformation $\textsl{x}(t)\rightarrow
r^m(t)=e^{-im\textsl{x}(t)}$ in Eq.~\eqref{F6} one obtains a
stochastic differential equation with a multiplicative noise term,
the averaging of which yields a system of differential-recurrence
relations for the moments $\langle r^m \rangle=\langle
e^{-im\textsl{x}} \rangle $ (as described in detail in Ref.~[14]),
viz.,
\begin{equation}
    \label{F7}
    \begin {array}{ll}
    \hat{\tau}d\langle r^m \rangle(t)/dt + [m^2/g + im \hat{F}_{Lx}]\langle r^m \rangle(t)= \\
    \\
    \qquad{} =(m/2)(\langle r^{m-1} \rangle(t)-\langle r^{m+1} \rangle(t)).\\
    \end{array}
\end{equation}

The main quantity of physical interest in our problem is the
average electric field, induced by the moving vortex system, which
is given by
\begin{equation}
    \label{F8}
    \langle \mathbf{E} \rangle=(n/c)\mathbf{B}\times \langle \mathbf{v} \rangle=n(B/c)(-\langle v_{y}\rangle\mathbf{x}+\langle v_{x}\rangle\mathbf{y}),
\end{equation}
where $\mathbf{x}$ and $\mathbf{y}$ are the unit vectors in the
$x$ and $y$ directions, respectively.

As follows from Eq.~\eqref{F5}
\begin{equation}
    \label{F9} \langle v_y \rangle=F_{Ly}/\eta+\delta\langle v_x\rangle
\end{equation}
and so for determination of $\langle \mathbf{E} \rangle$ from
Eq.~\eqref{F8} it is sufficient to calculate the $\langle v_x
\rangle$ from Eq.~\eqref{F6}. This calculation gives
\begin{equation}
    \label{F10}
    \langle v_x \rangle(t)=\frac{\Phi_{0}j_c}{c\eta D}[j^{dc}+j^{ac}\cos \omega t - \langle \sin \textsl{x}\rangle(t)],
\end{equation}
where
\begin{equation}
    \label{F11}
    \langle \sin \textsl{x} \rangle(t)=\frac{i}{2}[\langle r \rangle(t)-\langle r^{-1} \rangle(t)].
\end{equation}
In Eq.~(10) $j^{dc}\equiv n (j^{dc}_y+\delta j^{dc}_x)/j_c$,
$j^{ac}\equiv n (j^{ac}_y+\delta j^{ac}_x)/j_c$, and $j_c\equiv
cF_p/\Phi_0$.

Since we are only concerned with the stationary $ac$ response,
which is independent of the initial condition, one needs to
calculate the solution of Eq.~\eqref{F7} corresponding to the
stationary case. To accomplish this, one may seek all the $\langle
r^m \rangle(t)$ in the form
\begin{equation}
    \label{F12}
    \langle r^m\rangle(t)=\sum_{\substack{k=-\infty}}^{\substack{\infty}}F_k^m(\omega)e^{ik\omega t}.
\end{equation}
On substituting Eq.~\eqref{F12} into Eq.~\eqref{F7} we obtain
recurrence equations for the Fourier amplitudes $F_k^m(\omega)$,
i. e.,
\begin{equation}
    \label{F13}
    \begin {array}{ll}
    F_k^{m+1}(\omega) - F_k^{m-1}(\omega) + iz_{m,k}(\omega)F_k^m(\omega) + \\
    \\
    \qquad{}+i j^{ac}[F_{k-1}^m(\omega)+F_{k+1}^m(\omega)] =0,\\
    \end{array}
\end{equation}
where
\begin{equation}
    \label{F14}
    z_{m,k}(\omega)=2(j^{dc}+\omega \hat{\tau}k/m -im/g).
\end{equation}


\section{The solution of the problem in terms of matrix continued fractions}

The scalar five-term recurrence Eq.~\eqref{F13} can be transformed
into the two uncoupled matrix three-term recurrence relations
\begin{equation}
    \label{F15}
    \mathbf{Q}_m(\omega)\mathbf{C}_{m}(\omega)+\mathbf{C}_{m+1}(\omega)=\mathbf{C}_{m-1}(\omega),\quad(m=1,2,...)
\end{equation}
and
\begin{equation}
    \label{F16}
    -\mathbf{Q}_m^\ast(-\omega)\mathbf{C}_{-m}(\omega)+\mathbf{C}_{-m+1}(\omega)=\mathbf{C}_{-m-1}(\omega),\quad(m=1,2,...)
\end{equation}
where $\mathbf{Q}_m$ is a tridiagonal infinite matrix given by
\begin{equation}
    \label{F17}
    \mathbf{Q}_m(\omega)=i
        \begin{pmatrix}
            \ddots & \vdots & \vdots & \vdots & \ddots\\
            \cdots & z_{m,-2}(\omega) & j^{ac} & 0 & \cdots\\
            \cdots & j^{ac} & z_{m,-1}(\omega) & j^{ac} & 0\\
            \cdots & 0 & j^{ac} & z_{m,0}(\omega) & j^{ac}\\
            \cdots & \cdots & 0 & j^{ac} & z_{m,1}(\omega)\\
            \cdots & \cdots & \cdots & 0 & j^{ac}\\
            \ddots & \vdots & \vdots & \vdots & \ddots
        \end{pmatrix},
\end{equation}
(the asterisk denotes the complex conjugate) and the infinite
column vectors $\mathbf{C}_m(\omega)$ are defined as
\begin{equation}
    \label{F18}
    \begin{array}{ll}
        \mathbf{C}_m(\omega)=
            \begin{pmatrix}
                \vdots \\
                F^m_{-2}(\omega)\\
                F^m_{-1}(\omega)\\
                F^m_{0}(\omega)\\
                F^m_{1}(\omega)\\
                F^m_{2}(\omega)\\
                \vdots
            \end{pmatrix},
        \quad\textrm{for}\quad m=\pm1,\pm2,...\\
        \textrm{and}\quad
        \mathbf{C}_0=
            \begin{pmatrix}
                \vdots \\
                0\\
                0\\
                1\\
                0\\
                0\\
                \vdots
            \end{pmatrix},
        \quad\textrm{for}\quad m=0.
    \end{array}
\end{equation}

Thus, in order to calculate $\langle \sin \textsl{x} \rangle(t)$
in Eq.~\eqref{F11}, we need to evaluate $\mathbf{C}_1(\omega)$ and
$\mathbf{C}_{-1}(\omega)$, which contain all the Fourier
amplitudes of $\langle r \rangle(t)$ and $\langle r^{-1}
\rangle(t)$. Equation~\eqref{F15} can be solved for $\mathbf{C}_1$
in terms of matrix continued fractions$^{14}$, viz.,
\begin{equation}
    \label{F19}
    \mathbf{C}_1(\omega)=\cfrac{\mathbf{I}}{\mathbf{Q}_1(\omega)+\cfrac{\mathbf{I}}
    {\mathbf{Q}_2(\omega)+\cfrac{\mathbf{I}}{\mathbf{Q}_3(\omega)+...}}}\mathbf{C}_0,
\end{equation}
where the fraction lines designate the matrix inversions and
$\mathbf{I}$ is the identity matrix of infinite dimension. Having
determined $\mathbf{C}_1(\omega)$, it is not necessary to solve
Eq.~\eqref{F16}, as all the components of the column vector
$\mathbf{C}_{-1}(\omega)$ can be obtained from Eq.~\eqref{F19}, on
noting that
\begin{equation}
    \label{F20}
    F_0^1(\omega)=F_0^{-1\ast}(\omega) \quad \textrm{and}\quad F_k^{-1}(\omega)=F_{-k}^{1\ast}(\omega).
\end{equation}

Following the solutions of Eq.~\eqref{F19} and using
relations~\eqref{F20}, we can find the dimensionless average
pinning force $\langle F_{px}\rangle (t)$ (see
Eqs.~\eqref{F3}-\eqref{F6}, \eqref{F11}~and~\eqref{F12}) which is
the main anisotropic nonlinear (due to a dependence on the $ac$
and $dc$ current input) component of the theory under discussion
\begin{equation}
    \label{F21}
    \langle \hat{F}_{px}\rangle (t)=-\langle \sin\textsl{x}\rangle (t)=
    \sum_{\substack{k=0}}^{\substack{\infty}}\textrm{Im}(\psi_ke^{ik\omega t}),
\end{equation}
where $\psi_0\equiv F_0^1(\omega)$ and for $k\geq1$ we have
$\psi_k\equiv F_k^1(\omega)-F_k^{-1}(\omega)$.

In fact, Eq.~\eqref{F21} is the expansion of the stationary
time-dependent (and independent of the initial conditions) average
pinning force $\langle \hat{F}_{px}\rangle (t)$ into three parts
\begin{equation}
    \label{F22}
    \langle \hat{F}_{px}\rangle (t)=\langle \hat{F}_{px}\rangle_0^{\omega}+
    \langle \hat{F}_{px}\rangle_{t1}+\langle \hat{F}_{px}\rangle_t^{k>1}.
\end{equation}

In Eq.~\eqref{F22} $\langle
\hat{F}_{px}\rangle_0^{\omega}\equiv-\langle\sin\textsl{x}\rangle_0^{\omega}
= \textrm{Im}\psi_0$ is the time independent (but frequency
dependent) \textit{static} average pinning force, which will be
used for the derivation of the $dc$ magnetoresistivity tensor
$\hat{\rho}_0^{\omega}$; $\langle \hat{F}_{px}\rangle_{t1}\equiv
-\langle\sin\textsl{x}\rangle_{t1}=\textrm{Im}(\psi_1e^{i\omega
t})$ is the time-dependent \textit{dynamic} average pinning force
with a frequency $\omega$ of the $ac$ current input, which is
responsible for the nonlinear impedance $Z_1(\omega)$; $\langle
\hat{F}_{px}\rangle_t^{k>1}\equiv
-\langle\sin\textsl{x}\rangle_t^{k>1}=\textrm{Im}(\psi_ke^{ik\omega
t})$ describes a contribution of the \textit{harmonics} with $k>1$
into the dynamic average pinning force.


\section{$\omega$-dependent dc current magnetoresistivity
response}

\subsection{The nonlinear DC resistivity and conductivity tensors}

In order to proceed with these calculations we first express (see
Eq.~\eqref{F8}) the time independent part of $\langle E_y
\rangle(t)=(nB/c)\langle v_x\rangle(t)$ as
\begin{equation}
    \begin{array}{ll}
    \label{F23}
    \langle E_y \rangle_0^{\omega}=(nB/c)\langle v_x\rangle_0^{\omega}\equiv
    (n\rho_f/D)(j^{dc}-\langle \sin \textsl{x}\rangle_0^{\omega})=\\
    \\
    \qquad \qquad\qquad=(\rho_f/D)\nu_0^{\omega}(j^{dc}_y+\delta j^{dc}_x),
    \end{array}
\end{equation}
where
\begin{equation}
    \label{F24} \nu_0^{\omega}\equiv1-\langle \sin \textsl{x}\rangle_0^{\omega}/j^{dc}
    =1+\langle \hat{F}_{px}\rangle_0^{\omega}/j_{dc}.
\end{equation}

In Eq.~\eqref{F23} $\rho_f\equiv B\Phi_0/\eta c^2$ is the
flux-flow resistivity  and  the $\nu_0^{\omega}$ can be considered
as the $(\omega, j^{dc}, j^{ac}, T)$-dependent effective mobility
of the vortex  under the influence of the dimensionless
generalized moving force $\hat{F}_{Lx}^{dc}=j^{dc}$ in the $x$
direction. In the absence of the $ac$ current (see below Eq.(35))
the $\nu_0^{\omega}$ coincides with the probability of vortex
hopping over the pinning potential barrier$^3$.

From Eq.~\eqref{F24} follows \textit{another} physical
interpretation of the $\nu_0^{\omega}$ function, which has a close
relationship to the average pinning force $\langle
\hat{F}_{px}\rangle_0^{\omega}$ acting on the vortex. Actually, it
is evident from Eq.~\eqref{F24} that the $\langle
\hat{F}_{px}\rangle_0^{\omega}$ is connected to the
$\nu_0^{\omega}$ function in a simple way,
\begin{equation}
    \label{F25}
    \langle \hat{F}_{px}\rangle_0^{\omega}=-\hat{F}_{Lx}^{dc}(1-\nu_0^{\omega}).
\end{equation}

Then it is easy to show that
\begin{equation}
    \label{F26}
    \langle E_x\rangle_0^{\omega}=(\rho_f/D)[j^{dc}_x(1+\delta^2(1-\nu_0^{\omega}))-\delta j^{dc}_y].
\end{equation}

From Eqs.~\eqref{F23} and~\eqref{F26} we find the
$(\omega,j^{dc},j^{ac}, T)$-dependent magnetoresistivity tensor
for the $dc$-measured nonlinear law $\langle \mathbf{E}
_0^{\omega}(\omega) \rangle=\hat{\rho}_0^{\omega}\mathbf{j}^{dc}$
as
\begin{equation}
    \label{F27}
    \hat{\rho}_0^{\omega}=
        \begin{pmatrix}
            \rho_{xx}^{dc} & \rho_{xy}^{dc} \\
            \rho_{yx}^{dc} & \rho_{yy}^{dc}
        \end{pmatrix}
    =\frac{\rho_f}{D}
        \begin{pmatrix}
            1+\delta^2(1-\nu_0^{\omega}) & -\delta\nu_0^{\omega} \\
            \delta\nu_0^{\omega} & \nu_0^{\omega}
        \end{pmatrix}.
\end{equation}

The $dc$ conductivity tensor $\hat{\sigma}_0^{\omega}$, which is
the inverse tensor to $\hat{\rho}_0^{\omega}$, has the form
\begin{equation}
    \label{F28}
    \hat{\sigma}_0^{\omega}=
        \begin{pmatrix}
            \sigma_{xx}^{dc} & \sigma_{xy}^{dc} \\
            \sigma_{yx}^{dc} & \sigma_{yy}^{dc}
        \end{pmatrix}
    =\frac{1}{\rho_f}
        \begin{pmatrix}
            1 & \delta\\
            -\delta & (D/\nu_0^{\omega})-\delta^2
        \end{pmatrix}.
\end{equation}

We see from Eqs.~\eqref{F27} and~\eqref{F28} that the off-diagonal
components of the $\hat{\rho}_0^{\omega}$ and
$\hat{\sigma}_0^{\omega}$ tensors satisfy the Onsager relation
($\rho_{xy}=-\rho_{yx}$ in the general nonlinear case and
$\sigma_{xy}=-\sigma_{yx}$). All the components of the
$\hat{\rho}_0^{\omega}$ tensor and one of the diagonal components
of the $\hat{\sigma}_0^{\omega}$ tensor are function of the
current density $j^{dc}$, $j^{ac}$, and $\omega$ through the
external force $\hat{F}_{Lx}$, the temperature $T$, the angle
$\alpha$, and the dimensionless Hall parameter $\delta=n\epsilon$.
It is important, however, to stress that the off-diagonal
components of the $\hat{\sigma}_0^{\omega}$ are not influenced by
a presence of the pinning potential barriers.

\begin{figure}[t]
   \epsfig{file=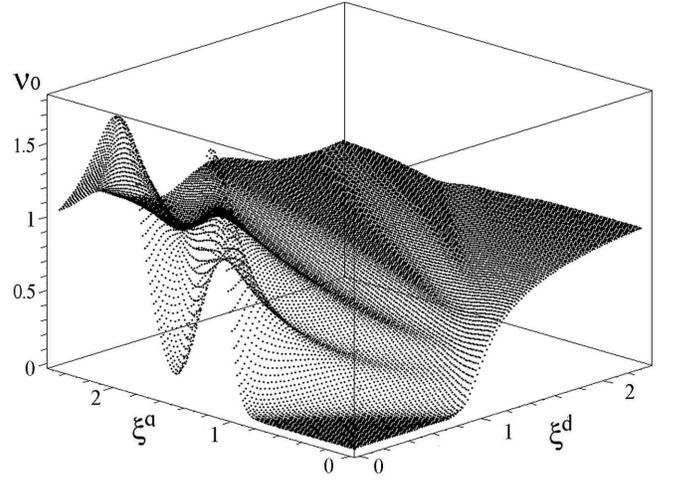,width=0.48\textwidth}
    \caption{The dimensionless function $\nu_0^{\omega}(\xi^d, \xi^a)$
    numerically obtained from Eq.~\eqref{F24}  at $g=30$ and $\Omega=0.2$.} \label{fig2}
\end{figure}

In conclusion of this subsection let us consider the limiting case
$j^{ac}=0$, i. e. we should derive a \textit{static}
current-voltage characteristic (CVC). In this case we have from
Eq.~\eqref{F7} that
\begin{equation}
    \label{D29}
    (m + igj^{dc})\langle r^m\rangle_0 = (g/2)(\langle r^{m-1}\rangle_0 - \langle
    r^{m+1}\rangle_0),
\end{equation}
where the subscript $"0"$ denotes the statistical average in the
absence of the $\textit{ac}$ current. In order to solve
Eq.~\eqref{D29} we introduce, following the calculation of
Risken\cite{RISK}, the quantity $S_m=\langle r^m\rangle_0/\langle
r^{m-1}\rangle_0$ which satisfies next equation
\begin{equation}
    \label{D30}
    (m + igj^{dc})S_m = (g/2)(1 - S_mS_{m+1}).
\end{equation}
The solution of Eq.~\eqref{D30} (see details in Ref.~[14]) can be
expressed in terms of the modified Bessel functions $I_{\nu}(z)$
of the first kind of order $\nu$ (where $\nu$ may be a complex
number~\cite{Watson}) as
\begin{equation}
    \label{D31}
    S_m = I_{m+\mu}(g)/I_{m-\mu}(g),
\end{equation}
where $\mu\equiv igj^{dc}$. Taking into account Eqs.~\eqref{F24},
\eqref{D31}, and the relation $S_1=\langle r\rangle_0=\langle \cos
\textsl{x} \rangle_0 - i \langle\sin \textsl{x} \rangle_0$ we
conclude that $\langle \hat{F}_{px}\rangle_0 = \textrm{Im}S_1 =
\textrm{Im}[I_{1+\mu}(g)/I_{\mu}(g)]$ and
\begin{equation}
    \begin{array}{ccr}
    \label{D32}
    \nu_0\equiv\nu_0^{\omega}(j^{ac}=0, \omega=0) = 1 +
    \textrm{Im}S_1/j^{dc}=\\
    \\
    = 1+ \textrm{Im}[I_{1+\mu}(g)/I_{\mu}(g)]/j^{dc}.\\
    \end{array}
\end{equation}
Note that Eq.~\eqref{D32} gives a more simple analytical
expression for the $\nu_0$ function which was presented in
Ref.~[2] on the basis of a Fokker-Planck approach, namely
\begin{equation}
    \label{D33}
    \nu^{-1}_0(F) = \frac{F}{1 - e^{-F}}\int_{\substack{0}}^{\substack{1}} due^{-Fu}I_0(2g\sin\pi u)
\end{equation}
with $F\equiv 2\pi g j^{dc}$, where $g=U_p/2T$ is the
dimensionless inverse temperature. In Fig.~2 we plotted
$\nu_0^{\omega}(\xi^d,\xi^a)$ graphs at g=30 which demonstrate in
the limit of $\xi^a=0$ the $\xi^d$-dependence for probability of
vortex hopping over the tilted cosine pinning potential barrier;
here $\xi^d$ and $\xi^a$ are the dimensionless $dc$ and maximal
$ac$ current density magnitudes (in $j_c$ units), respectively
($\xi^d\equiv j^d/j_c, \xi^a\equiv j^a/j_c$).


\subsection{Longitudinal and transverse DC resistivities}

The experimentally measurable resistive $dc$ responses refer to
coordinate system tied to the $dc$ current (see Fig.~1). The
longitudinal and transverse (with respect to the $dc$ current
direction) components of the electric field $E_{\parallel}^{dc}$
and $E_{\perp}^{dc}$, are related to $E_x^{dc}\equiv\langle E_x
\rangle_0^{\omega}$ and $E_y^{dc}\equiv\langle E_y
\rangle_0^{\omega}$ by the simple expressions
\begin{equation}
    \label{F29}
    \begin{array}{crr}
    E_{\parallel}^{dc}=E_x^{dc}\sin\alpha + E_y^{dc}\cos\alpha,\\
    \\
    E_{\perp}^{dc}= - E_x^{dc}\cos\alpha + E_y^{dc}\sin\alpha.\\
    \end{array}
\end{equation}

Then according to Eqs.~\eqref{F29}, the expressions for the
experimentally observable longitudinal and transverse (with
respect to the $\mathbf{j}^{dc}$ direction) magnetoresistivities
$\rho_{\parallel}^{dc}=E_{\parallel}^{dc}/j^d$ and
$\rho_{\perp}^{dc}=E_{\perp}^{dc}/j^d$ (where $j^d$ is the $dc$
current density $(j^d)^2=(j_x^{dc})^2 + (j_y^{dc})^2$) have the
form
\begin{equation}
    \label{F30}
    \begin{array}{crr}
    \rho_{\parallel}^{dc}=\rho_{xx}^{dc}\sin^2\alpha + \rho_{yy}^{dc}\cos^2\alpha,\\
    \\
    \rho_{\perp}^{dc}= \rho_{yx}^{dc}\sin^2\alpha -\rho_{xy}^{dc}\cos^2\alpha + (\rho_{yy}^{dc}-\rho_{xx}^{dc})\sin\alpha\cos\alpha.\\
    \end{array}
\end{equation}

Note, however, that the magnitudes of the $\rho_{\parallel}^{dc}$
and $\rho_{\perp}^{dc}$, given by Eqs.~\eqref{F30} and applied to
the $dc$ current responses, in general, depend on the direction of
the external magnetic field $\mathbf{B}$ along $z$ axis due to the
$\delta=n\epsilon$ dependence of the $\nu_0^{\omega}$ function
(see Eq.~\eqref{F24}). In order to consider only $n$-independent
magnitudes of the $\rho_{\parallel}^{dc}$ and $\rho_{\perp}^{dc}$
responses we should introduce the even $(+)$ and odd $(-)$
magnetoresistivities with respect to magnetic field reversal
($\rho^{dc\pm}(n) \equiv [\rho^{dc}(n) \pm \rho^{dc}(-n)]/2$) for
longitudinal and transverse dimensional magnetoresistivities,
which in view of Eqs.~\eqref{F30} have the form
\begin{equation}
    \label{F31}
    \begin{array}{ccr}
    \rho_{\parallel}^{dc\pm}=(\rho_f/D)[(\cos^2\alpha - \delta^2\sin^2\alpha)\nu_0^{\omega\pm}+\\
    \\
    +D(1\pm1)\sin^2\alpha/2],\\
    \end{array}
\end{equation}
\begin{equation}
    \label{F32}
    \begin{array}{ccr}
    \rho_{\perp}^{dc\pm}=(\rho_f/D)[D\nu_0^{\omega\pm} \sin\alpha\cos\alpha + \delta\nu_0^{\omega\mp}-\\
    \\
    -D(1\pm1)\sin\alpha\cos\alpha/2],\\
    \end{array}
\end{equation}
where $\nu_0^{\omega\pm}(n)=[\nu_0^{\omega}(n) \pm
\nu_0^{\omega}(-n)]/2$ are the even and odd components relative to
the magnetic field inversion of the function $\nu_0^{\omega}(n)$.
In the $E_{\parallel,\perp}^{dc+}(j)$ dependences, which follow
from Eqs.~\eqref{F31} and~\eqref{F32}, the nonlinear and linear
(nonzero only for $\rho_\parallel^{dc+}$ and $\rho_\perp^{dc+}$)
terms separate out in a natural way. The physical reason for the
appearance of linear terms is that in the model under
consideration for $\alpha\neq0$ there is always the flux-flow
regime of vortex motion along the channels of the PPP.

It follows from Eqs.~\eqref{F31} and~\eqref{F32} that for
$\alpha\neq0, \pi/2$ the observed resistive response contains not
only the ordinary longitudinal $\rho_\parallel^{dc+}(\alpha)$ and
transverse $\rho_\perp^{dc-}(\alpha)$ magnetoresistivities, but
also (as in the absence of $ac$ current, see Ref.~[3]) two new
components, induced by the pinning anisotropy: an \textit{even
transverse} $\rho_\perp^{dc+}(\alpha)$ and the \textit{odd
longitudinal} component $\rho_\parallel^{dc-}(\alpha)$.

In the absence of $ac$ current
($\nu_0\equiv\nu_0^{\omega}(\xi^a=0, \omega=0)$) the physical
origin of the $\rho_\perp^{dc+}(\alpha)$ (which is independent of
$\epsilon$ at $\epsilon\ll1$) is related to the guided vortex
motion along the channels of the washboard pinning potential in
the TAFF regime. On the other hand, the
$\rho_\parallel^{dc-}(\alpha)$ component is proportional to the
odd component $\nu_0^-$ which is zero at $\epsilon=0$ and has a
maximum in the region of the nonlinear transition from the TAFF to
the FF regime at $\epsilon\neq0$ (see Figs.~6 and~7 in Ref.~[3]).
The $(j^{dc},~g)$ dependence of the odd transverse (Hall)
resistivity has contributions both from the even $\nu_0^+$ and
from the odd $\nu_0^-$ components of the $\nu_0(j^{dc},~g)$
function. Their relative magnitudes are determined by the angle
$\alpha$ and the dimensionless Hall constant $\epsilon$. Note that
as the odd longitudinal $\rho_\parallel^{dc-}$ and odd transverse
$\rho_\perp^{dc-}$ magnetoresistivities arise by virtue of the
Hall effect, their characteristic scale is proportional to
$\epsilon\ll1$.


\subsection{DC response in LT geometries}

In order to analyze the most simple forms of the
$\rho_{\parallel,\perp}^{dc\pm}$ equations given by
formulas~\eqref{F31} and~\eqref{F32} we introduce the $L$ and $T$
geometries (see Fig.~1), in which $\mathbf{j}\parallel\mathbf{x}$
(i.~e.~$\alpha=\pi/2$) and $\mathbf{j}\perp\mathbf{x}$
(i.~e.~$\alpha=0$ and $\mathbf{j}\parallel\mathbf{y}$),
respectively. It follows from Eqs.~\eqref{F31} and~\eqref{F32}
that the longitudinal $\rho_{\parallel}^{dc-}$ and transverse
$\rho_{\perp}^{dc+}$ resistivity for a superconductor with
uniaxial pinning anisotropy in $LT$ geometries vanish (i.~e.
$\rho_{\parallel}^{dc-} = \rho_{\perp}^{dc+} = 0$) and we obtain
\begin{equation}
    \label{F33}
    \rho_{\parallel,T}^{dc+}=\nu_{0,T}^{\omega}/D, \quad
    \rho_{\perp,T}^{dc-}=n\epsilon\nu_{0,T}^{\omega}/D,
\end{equation}
\begin{equation}
    \label{F34}
    \rho_{\parallel,L}^{dc+}=1-\epsilon^2\nu_{0,L}^{\omega}/D, \quad
    \rho_{\perp,L}^{dc-}=n\epsilon\nu_{0,L}^{\omega}/D.
\end{equation}

Here $\nu_{0,T}^{\omega}\equiv\nu_0^{\omega}(j_T^{dc}, j_T^{ac},
\omega, g)$, $\nu_{0,L}^{\omega}\equiv\nu_0^{\omega}(j_L^{dc},
j_L^{ac}, \omega, g)$, $j_T^{dc}\equiv n\xi^d$, $j_T^{ac}\equiv
n\xi^a$, $j_L^{dc}\equiv \epsilon \xi^d$, $j_L^{ac}\equiv \epsilon
\xi^a$.

If we neglect the Hall terms in Eqs.~\eqref{F33} and~\eqref{F34},
then in the absence of an $ac$ current in the $L$ geometry vortex
motion takes place along the channels of the washboard PPP (the
\emph{guiding} effect), and in  the $T$ geometry - transverse to
the washboard channels (the \emph{slipping} effect). In the $L$
geometry the critical current is equal to zero since the FF regime
is realized for the guided vortex motion along the PPP channels.
In the $T$ geometry, i.~e.~for vortex motion transverse to the
channels, a pronounced nonlinear regime is realized for $g\gg 1$,
the onset of which corresponds to the crossover point
$j^d=j_{cr}$, and for $g\gg 1$ we have $j_{cr}=j_c$, where $j_c$
is the critical current. The longitudinal even
$\rho_{\parallel,T}^{dc+}$ and transverse odd
$\rho_{\perp,T}^{dc-}$ resistivities are proportional to the even
function $\nu_{0,T}^+(\xi^d, g)$. In the limit
$j^d,~g\rightarrow0$ to within terms, proportional to
$\epsilon^2\ll1$, we have $\rho_{\parallel,T}^{dc+}=1$ and
$\rho_{\perp,T}^{dc-}=n\epsilon$. The main contribution to the
$\rho_{\parallel,L}^{dc+}$ which is equal to 1 with the same
accuracy, is due to the guided vortex motion along the washboard
channels where the pinning is absent. The magnitude of
$\rho_{\perp, L}^{dc-}$ resistivity is described by the Magnus
force $\epsilon\xi^d$ which is vanishingly small for a small Hall
effect for realistically achievable currents $j^d\ll j_c/\epsilon$
and the velocity component $\langle v_x\rangle$ is suppressed, the
resistivity $\rho_{\perp,L}^{dc-}$ depends mainly only on the
temperature. For $g \gg 1$ the $\rho_{\perp,L}^{dc-}$ is so small
that it cannot be measured $(\rho_{\perp,L}^{dc}=0$ in the limit
$g\gg 1$ since $\epsilon\xi^d<1)$, and for $g\sim 1/2$ it
approaches the value of the Hall constant, $\epsilon$ (to within
terms proportional to $\epsilon^2\ll1$).

It is worth noticing that simple Eqs.~\eqref{F33} in the $T$
geometry allow one to extract from the $(\xi^d,~\xi^a,~\omega,~g)$
dependences of the measured resistivities
$\rho_{\parallel,T}^{dc+}$ and $\rho_{\perp,T}^{dc-}$ the
dimensionless Hall constant $\epsilon$ and the main nonlinear
component of the model under discussion $\nu_{0,T}^{\omega}$. The
latter in the absence of $ac$ current, i.~e.~$\nu_{0,T}$, can be
used for the prediction of the $\alpha$-dependent
$\rho_{\parallel,\perp}^{dc\pm}$ resistivities given by
Eqs.~\eqref{F31} and~\eqref{F32} in the case of $\epsilon\ll1$.


\subsection{Guiding of vortices and the Hall effect in nonlinear DC+AC regimes}

After derivation of Eqs.~\eqref{F31} and ~\eqref{F32} let us
proceed now to a more detailed treatment of the $dc$ vortex
dynamics and the resistive properties associated with them in the
presence of an $ac$ current. For simplicity we will neglect the
usually small Hall effect, i.~e. we take $\epsilon=0$. As a
consequence, the nondiagonal components of the $dc$
magnetoresistivity tensor (see Eq.~\eqref{F27}) vanish
($\rho_{xy}^{dc}=\rho_{yx}^{dc}=0$). Neglecting the Hall effect,
the formulas for the experimentally observed longitudinal
$\rho_\parallel^{dc}$ and transverse $\rho_{\perp}^{dc}$
resistivities relative to the $dc+ac$ current can be represented
as
\begin{equation}
    \label{G1}
    \rho_{\parallel}^{dc}=\rho_f(\nu_0^{\omega}\cos^2\alpha + \sin^2\alpha), \quad
\end{equation}
\begin{equation}
    \label{G2}
    \begin{array}{crr}
    \rho_{\perp}^{dc}=\rho_f[(\nu_0^{\omega}-1)\sin\alpha\cos\alpha]=\\
    \\
    =\rho_fj^{dc}\langle \hat{F}_{px} \rangle^{\omega}_0\sin\alpha\cos\alpha.
    \end{array}
\end{equation}

Therefore, as was pointed out in Ref.~[3], even in the absence of
$ac$ current, under certain conditions in the $dc$ current and
temperature dependences of the $\rho^{dc}_{\parallel}$ and
$\rho^{dc}_{\perp}$ a pronounced nonlinearity appears in the
vortex dynamics and a nonlinear guiding effect may be observed in
both the inverse temperature $g$ and the current density
$j^{dc}=n\xi^d_y$. As a consequence of the even parity of
$\nu_0^{\omega}$ in $j^{dc}$ and $j^{ac}=n\xi^a_y$ (see
Eqs.~\eqref{F17}-\eqref{F20} and~\eqref{F24}) the
magnetoresistivities $\rho^{dc}_{\parallel}$ and
$\rho^{dc}_{\perp}$ are even in the magnetic field reversal, as
they should be neglecting the Hall effect.

As was shown in Ref.~[3], the specifics of anisotropic pinning
consist in the noncoincidence of the directions of the external
motive Lorentz force $\mathbf{F}_L$ acting on the vortex, and its
velocity $\langle \mathbf{v} \rangle$ (for isotropic pinning
$\mathbf{F}_L\parallel \langle \mathbf{v} \rangle$ if we neglect
the Hall effect). The anisotropy of the pinning viscosity (which
can be defined as the inverse vortex mobility
$[\langle\nu\rangle_0^{\omega}]^{-1}$) along and transverse to the
PPP channels leads to the result that for those values of
($j^{dc}, g, \alpha$) for which the component of the vortex
velocity perpendicular to the PPP channels, $\langle v_x
\rangle_0^{\omega}$, is suppressed, a tendency appears toward a
substantial prevalence of guided vortex motion along PPP channels
(the guiding effect) over motion transverse to the channels (the
slipping effect). In the experiment, the function
\begin{equation}
\label{G3}
    \cot\beta=-\frac{\rho_{\perp}^{dc}}{\rho_{\parallel}^{dc}}=
    \frac{1-\nu^{\omega}_0(\xi^d_y,g,\xi^a_y)}{\tan\alpha+\nu_0^{\omega}(\xi^d_y,g,\xi^a_y)\cot\alpha}
\end{equation}
is used to describe the guiding effect, where $\beta$ is the angle
between the average vortex velocity vector $\langle \mathbf{v}
\rangle$ and the current density vector $\mathbf{j}^{dc}$ (see
Fig.~1). The guiding effect is more stronger when the difference
in directions of $\mathbf{F}_L$ and $\langle \mathbf{v} \rangle$
is larger, i.~e., the smaller is the angle $\beta$. Let us
consider the current and temperature dependence of
$\cot\beta(\xi^d_y,g,\xi^a_y)$ for fixed values of the angle
$\alpha\neq0, \pi/2$. In the temperature region corresponding to
the TAFF regime, we have $\beta\approx\alpha$ and, consequently,
at low currents guiding arises. At large currents ($\xi_y^d\gg1$),
where for vortex motion transverse to the PPP channels the FF
regime is set up, i.~e. the vortex dynamics becomes isotropic and
we have $\langle \mathbf{v} \rangle \parallel \mathbf{F}_L$ for
arbitrary value of the angle $\alpha$.

Let us now analyze the $dc$ magnetoresistivity dependences
$\rho^{dc\pm}_{\parallel}$ and $\rho^{dc\pm}_{\perp}$, given by
Eqs.~\eqref{F31} and \eqref{F32}, with allowance for the small
Hall effect. In this case, the expressions for
$\rho^{dc\pm}_{\parallel,\perp}$, out to terms of order
$\epsilon^2\ll1$, have the form
\begin{equation}
\label{G4}
    \begin{array}{crr}
    \rho^{dc+}_{\parallel}=\rho_f(\nu_0^{\omega+}\cos^2\alpha+\sin^2\alpha),\\
    \\
    \rho^{dc+}_{\perp}=\rho_f(\nu_0^{\omega+}-1)\sin\alpha\cos\alpha,
    \end{array}
\end{equation}
\begin{equation}
\label{G5}
    \begin{array}{crr}
    \rho^{dc-}_{\parallel}=\rho_f\nu_0^{\omega-}\cos^2\alpha,\\
    \\
    \rho^{dc-}_{\perp}=\rho_f(\delta\nu_0^{\omega+}+\nu_0^{\omega-}\sin\alpha\cos\alpha),
    \end{array}
\end{equation}

Here $\nu_0^{\omega\pm}$ are obtained from relations
\begin{equation}
\label{G6}
    \nu_0^{\omega}=1+ \langle \hat{F}_{px} \rangle_0^{\omega}/j^{dc}\equiv1 + \textrm{Im}G_0,
\end{equation}
where $G_0\equiv\psi_0(j^{dc}, j^{ac}, \omega, g)/j^{dc}$ and
\begin{equation}
\label{G7}
    \nu_0^{\omega+}=1+ \textrm{Im}G_0^+,\quad\nu_0^{\omega-}=\textrm{Im}G_0^-.
\end{equation}

In the limit of a small Hall effect $(\epsilon\ll1)$ the
expressions for even and odd components of $\nu_0^{\omega}$ (in
terms of $G_0^{\pm}$) in the linear approximation in the parameter
$\epsilon\tan\alpha\ll1$ are equal respectively to
\begin{equation}
\label{G8}
    \begin{array}{crr}
    G_0^+=G_0(n\xi_y^d, n\xi_y^a) + nR_d^+\delta\tan\alpha\\
    \\
    G_0^-=(nR_d^- - G_0^+)\delta\tan\alpha,
    \end{array}
\end{equation}
\begin{equation}
\label{G9}
    R_d\equiv[\partial\psi_0/\partial\xi_y^d + (j^a/j^d)(\partial\psi_0/\partial\xi_y^a)],
\end{equation}
where $\psi_0=\psi_0(n\xi_y^d, n\xi_y^a)$, $j^d$ and $j^a$ are
$dc$ and $ac$ current density values, and $R_d^+$, $R_d^-$ are
even and odd parts of the $R_d$, respectively.

As follows from Eq.~\eqref{G4} and~\eqref{G5}, the behavior of the
$dc$ current and temperature dependence of
$\rho^{dc\pm}_{\parallel,\perp}$ is completely determined by the
$(j^d, g|j^a, \omega)$-behavior of the $\nu_0^{\omega\pm}$
dependences. If $(j^a/j^d)\ll1$, i.~e. the influence of the $ac$
current on the $dc$ response can be considered as a small, the
linear limit for $E^{dc\pm}_{\parallel,\perp}(j^{dc})$
dependences, following from Eqs.~\eqref{G4} and ~\eqref{G5}, is
realized in that region of $j^d$ and $g$, where
$\nu_0^{\omega+}=const$ and $\nu_0^{\omega-}=0$, while the region
of nonlinearity of $\rho^{dc\pm}_{\parallel,\perp}(j^d, g|j^a,
\omega)$ dependences corresponds to those $j^d$ and $g$ intervals,
where the dependences $\nu_0^{\omega\pm}(\xi^d_y|g)$ and
$\nu_0^{\omega\pm}(g|\xi^d_y)$ are nonlinear. Note, that the
nonlinearity in the temperature dependences
$\rho^{dc\pm}_{\parallel,\perp}(g|\xi^d_y)$ can be observed not
only at small currents (in the TAFF regime), but even at large
currents $\xi^d>1$ in the case when $\xi_y^d<1$, where this latter
relation depends on the magnitude of the angle $\alpha$ (for
$\cos\alpha<1/\xi^d$ we have $\xi_y^d<1$ and
$\nu_0^{\omega}(g\gg1)=0$). Thus, the linearity or nonlinearity of
the dependences $\rho^{dc\pm}_{\parallel,\perp}(g)$ at $dc$
currents larger than unity depends on the magnitude of the
angle~$\alpha$.


\subsection{Shapiro steps and adiabatic DC response}

Before a discussion about the influence of the $ac$ current on the
current-voltage characteristic (CVC) of the model under discussion
it is instructive to consider first a simple physical picture of
the vortex motion in a \textit{tilted} (due a presence of the
dimensionless $dc$ driving force $0<\xi^d<\infty$) washboard
planar pinning potential (PPP) under the influence of the
\textit{effective} dimensionless driving force $\hat{f}=
\hat{F}_{px}+\hat{F}_{Lx}=-\sin\textsl{x}+\xi^d$.

If the temperature is zero, the vortex is at rest with $\xi^d=0$
at the bottom of the potential well of the PPP. When the PPP is
gradually lowered by increasing $\xi^d$, then for $0<\xi^d<1$
appears an asymmetry of the left-side and right-side potential
barriers for a given potential well, and in this range of $\xi^d$
an effective force $\hat{f}$ changes its sign periodically. With
gradual $\xi^d$-increasing there will come a point where
$\xi^d=1$, and for $\xi^d>1$ the more lower right-side potential
barrier disappears, the effective motive force $\hat{f}$ becomes
everywhere along \textsl{x} positive and the vortex is in the
"running" state, periodically changing its velocity with a
dimensionless frequency $\omega_i=\sqrt{(\xi^d)^2-1}$. So the
static CVC of this periodic motion at $\xi^d>1$ is a result of
time-averaging of the stationary time-dependent solution of the
equation of motion $d\textsl{x}/d\tau=\hat{f}$ with
$\tau=t/\hat{\tau}$. Eventually, the probability of the vortex
overcoming the barriers of the PPP
$\nu_0\equiv\nu_0^{\omega}(\xi^a=0, \omega=0)$ at zero temperature
is
\begin{equation}
    \label{DC1}
    \nu_0=
    \left\{
        \begin{array}{crr}
        \, 0, \qquad\qquad \qquad\xi^d<1,\\
        \\
        \sqrt{1-(1/\xi^d)^2}, \quad \xi^d\geq1,\\
        \end{array}
    \right.
\end{equation}
i.~e. the $\nu_0(\xi^d>1)$ monotonically tends to unity with
$\xi^d$-increasing.

If the temperature is nonzero, a diffusion-like mode appears in
the vortex motion. At low temperatures ($g\gg1$) and $0<\xi^d<1$
the thermoactivated flux-flow (TAFF) regime of the vortex motion
occurs by means of the vortex hopping between neighboring
potential wells of the PPP. The intensity of these hops at low
temperatures is proportional to the $~\exp{[-g(1-\xi^d)]}$, i.~e.
strongly increases with $T$-increasing and $\xi^d$-increasing due
to the lowering of the right-side potential barriers at their
tilting. On the other hand, at $\xi^d$ just above the unity (when
the running mode is yet weak), the diffusion-like mode can
strongly increase the average vortex velocity even at relatively
low temperature due to a strong enhancement of the effective
diffusion coefficient of an overdamped Brownian particle in a
tilted PPP near the critical tilt~\cite{RIEN} at $\xi^d=1$ (see
below subsection V.~H.).

\begin{figure}[t]
    \epsfig{file=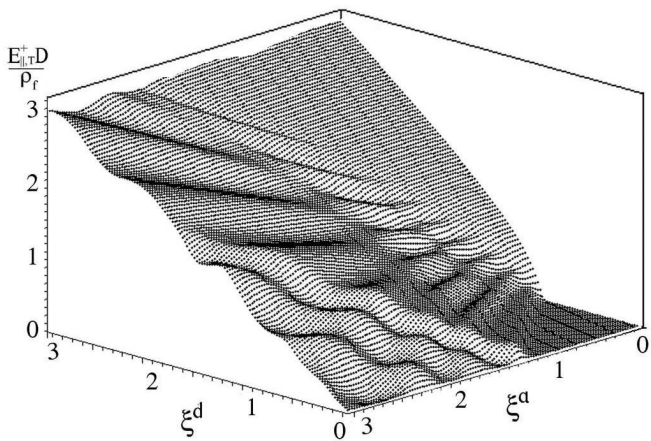,width=0.48\textwidth}
    \caption{The longitudinal CVC $E^{dc+}_{\parallel,\textrm{T}}(\xi^d, \xi^a),
    \Omega=0.2, g=100$, showing the Shapiro steps.} \label{fig2}
\end{figure}

Now we consider the influence of a small ($\xi^a\ll1$) $ac$
current density with a frequency $\omega$ on the CVC in the limit
of very small temperatures ($g\ggg1$). In this case the physics of
the $dc$ response is quite different depending on the $\xi^d$
value with respect to the unity. If $\xi^d<1$, the vortex mainly
(excluding very rare hops to the neighboring wells) localized at
the bottom of the potential well where it experiences a small
$\omega$-oscillations. The averaging of the vortex motion over the
period of oscillations in this case cannot change the CVC which
existed in the absence of the $ac$-drive.

If, however, $\xi^d>1$, then the vortex is in a running state with
the internal frequency of oscillations
$\omega_i=\sqrt{(\xi^d)^2-1}$. If $\omega\neq\omega_i$, the CVC is
changed only in the second-oder perturbation approach in terms of
a small parameter $\xi^a\ll1$ (as it was shown for the analogous
resistively shunted Josephson junction problem~\cite{KANVER})
because the CVC is not changed in the linear approximation in this
case. However, for $\omega=\omega_i$ appears a problem of a
synchronization of the running vortex oscillations at the
$\omega_i$-frequency with the external driving frequency $\omega$.
As a result, the average (over period of oscillations) vortex
velocity is locked in with the $\omega$ in some interval of the
$dc$ current density $\xi^d$ even within the frame of the
first-order perturbation calculation. The width of this first
synchronization step (or the so-called "Shapiro step" in the
resistively shunted Josephson junction problem) has been found in
Ref.~[\cite{ASLAR}] and the calculation in the spirit of this
reference gives the boundaries of the $\xi^d$ where the step
occurs as
\begin{equation}
\label{DC2}
    (\xi^d_{\omega} - \xi^a/2\xi^d)<\xi^d<(\xi^d_{\omega} +
    \xi^a/2\xi^d).
\end{equation}
Here $\xi^d_{\omega}$ is the current density which gives
$\omega_i=\sqrt{(\xi^d_{\omega})^2-1}=\omega$, i.~e.
$\xi^d_{\omega}=\sqrt{1+\omega^2}$. Then the size of the first
Shapiro step on the CVC is $\xi^a/\xi^d_{\omega}$. In higher
approximations (in terms of ($\xi^a)^m$), where $m$ runs through
all of the integers) the Shapiro steps on the CVC appear at the
frequencies $\Omega=m\omega$ and $\Omega_i=m\omega_i$. The width
of the $m$-th step at $\xi^a\rightarrow0$ is proportional to
$(\xi^a)^m$, i.~e. strongly decreases with $m$
increasing~\cite{LIHUL}.

In Fig.~3 we plot the longitudinal CVC
$E^{dc+}_{\parallel,\textrm{T}}(\xi^d, \xi^a)$ showing the Shapiro
steps. The plot in Fig.~3 looks like the similar curves discussed
earlier~\cite{VAN} for the CVC of the microwave driven resistively
shunted Josephson junction model at $T=0$ where the overall shape
of the CVC and different behaviour of the two types of the Shapiro
steps in adiabatic limit was explained. Our graph, in comparison
with the curves of Ref.~\cite{VAN}, is smoothed due to the
influence of a finite temperature. The longitudinal CVC
$E^{dc+}_{\parallel,\textrm{T}}(\xi^d, \xi^a)$-dependence
demonstrate several main features. First, in the presence of the
microwave current the $dc$ critical current $\xi^d_c(\xi^a)$ is a
decreasing function of the $ac$ driving. The physical reason for
such behaviour lies in the replacement of the $dc$ critical
current by the total $dc+ac$ critical current. Second, with
gradual $\xi^a$-increasing the zero-voltage step reduces to zero
and all other steps appear. Such steps are common because they do
not oscillate and spread over a $dc$-current range about twice the
critical current $2\xi^a_c$. These steps are the steps of the
first kind and they distort the CVC as like as relief bump with a
concave shift from the ohmic line. With further $\xi^a$-increasing
this relief bump shifts toward higher $\xi^d$-values. Below this
range the steps of the second kind appear. These microwave
current-induced steps oscillate rapidly and stay closely along the
ohmic line~\cite{VAN} over a $dc$-current range
$\xi^d\leq\xi^a-1$.

To summarize, we can determine three $(\xi^d,\xi^a)$-ranges where
the CVC-behaviour is qualitatively different. For large $dc$ bias
current densities $\xi^a+1<\xi^d$ the CVC asymptotically
approaches the ohmic line without microwave induced steps. For an
intermediate $dc$ current range $\xi^a-1<\xi^d<\xi^a+1$ CVC curve
deviates from the ohmic line as a concave bump with the stable
steps. For lower $dc$ current range $\xi^d<\xi^a-1$ the steps
oscillate with microwave current along the ohmic line. With
gradual $\Omega$-increasing the size of the steps increases
whereas their number decreases.


\section{Nonlinear stationary ac response}

\subsection{Derivation of the impedance tensor}

Using Eq.~\eqref{F8} we determine nonlinear (in the amplitudes
$j^{ac}$, $j^{dc}$ and the frequency $\omega$) stationary $ac$
response as
\begin{equation}
    \label{F35}
    \langle \mathbf{E} \rangle _t\equiv\langle\langle\mathbf{E} \rangle (t)-\langle\mathbf{E}\rangle_0^{\omega}\rangle=(nB/c)[\langle v_x \rangle_t\mathbf{y}-\langle v_y \rangle_t \mathbf{x}],
\end{equation}
where $\langle \mathbf{E}\rangle_0^{\omega}=(nB/c)[-\langle v_y
\rangle_0^{\omega} \mathbf{x}+\langle v_x \rangle_0^{\omega}
\mathbf{y}]$ is time-independent part of $\langle \mathbf{E}
\rangle (t)$ (see also Eqs.~\eqref{F23} and \eqref{F26}), whereas
$\langle v_y \rangle_t$ and $\langle v_x \rangle_t$ are
time-dependent periodic parts of $\langle v_y \rangle (t)$ and
$\langle v_x \rangle (t)$ which to become zero after averaging
over a period $2\pi/\omega$ of the $ac$ cycle.

From Eqs.~\eqref{F10} and~\eqref{F35} we have
\begin{equation}
    \label{F36}
    \langle E_y \rangle _t=(n\rho_fj_c/D)\sum_{\substack{k=1}}^{\substack{\infty}}(j^{ac}) ^k \textrm{Re}\{Z_k(\omega)e^{ik\omega t}\},
\end{equation}
where
\begin{equation}
    \label{F37}
    \begin{array}{ccr}
    Z_k(\omega)=\delta_{1,k}-i\psi_k(\omega)/(j^{ac})^k, \\
    \\
    \psi_k(\omega)\equiv[F_k^1(\omega)-F_k^{-1}(\omega)],
    \end{array}
\end{equation}
and $\delta_{1,k}$ is Kronecker`s delta.

The dimensionless transformation coefficients $Z_k$ in
Eq.~\eqref{F36} have a physical meaning of the $k$-th harmonic
with frequency $\Omega_k\equiv k\omega$ in the $ac$ nonlinear
$\langle E_y \rangle _t$ response. Equation~\eqref{F37} for $k=1$
yields
\begin{equation}
    \label{F38}
    Z_1=1-i\psi_1/j^{ac},
\end{equation}
and using Eqs.~\eqref{F35} and~\eqref{F36} we can express the
nonlinear stationary $ac$ response on the $\omega$-frequency
$E_{y1}^{ac}$, in terms of the nonlinear impedance $Z_1$ as
\begin{equation}
    \label{F39}
    E_{y1}^{ac}=(\rho_f/D)(j^{ac}_y+\delta j^{ac}_x)\textrm{Re}\{Z_1e^{i\omega t}\}.
\end{equation}

If we put $Z_1\equiv\rho_1-i\zeta_1$, where $\rho_1$ and $\zeta_1$
are the dynamic resistivity and the reactivity, respectively, then
Eq.~\eqref{F39} acquires form
\begin{equation}
    \label{F40}
    E_{y1}^{ac}=(\rho_f/D)(j^{ac}_y+\delta j^{ac}_x)\sqrt{\rho_1^2+\zeta_1^2}\cos(\omega t - \varphi_1),
\end{equation}
where $\sqrt{\rho_1^2+\zeta_1^2}\equiv|Z_1|$ and
$\varphi_1=\arctan(\zeta_1/\rho_1)$ are the dimensionless
amplitude and phase of the $ac$ response on the
$j^{ac}_{Lx}=(j^{ac}_y+\delta j^{ac}_x) \cos \omega t$ input.

Similarly, using Eq.~\eqref{F9} and~\eqref{F35}, we can show that
\begin{equation}
    \label{F41}
    \langle E_x \rangle _t=\rho_fj^{ac}_x \cos\omega t - \delta  \langle E_y \rangle_t,
\end{equation}
and obtain $\omega$-frequency $ac$ response $E_{x1}^{ac}$ as
\begin{equation}
    \label{F42}
    E_{x1}^{ac}=(\rho_f/D)\textrm{Re}\{e^{i\omega t}[(D - \delta^2Z_1)j_x - \delta Z_1j_y)]\}.
\end{equation}
From Eqs.~\eqref{F39} and~\eqref{F42} follows that the complex
amplitudes of the electric field $\textsl{\textbf{E}}_1$ and the
current density $\textsl{\textbf{J}}=\textbf{j}e^{i\omega t}$ are
connected by the relation $\textsl{\textbf{E}}_1 =
\hat{Z}\textsl{\textbf{J}}$, where $\hat{Z}$ is the frequency and
$dc$ and $ac$ current amplitudes dependent impedance tensor
\begin{equation}
    \label{F43}
    \hat Z(\omega)=
        \begin{pmatrix}
            Z_{xx} & Z_{xy} \\
            Z_{yx} & Z_{yy}
        \end{pmatrix}
    =   \frac{\rho_f}{D}
        \begin{pmatrix}
            D - \delta^2 Z_1 & -\delta Z_1\\
            \delta Z_1 & Z_1
        \end{pmatrix}.
\end{equation}

It is relevant to remark the similarity of Eq.~\eqref{F27} and
Eq.~\eqref{F43} from which follows that for the $ac$ response
$Z_1$ plays the same role as $\nu_0^{\omega}$ for the $dc$
response.

However, the connection between $Z_1$ and the dynamical average
pinning force $\langle \hat{F}_{px} \rangle_{t1}$ is more complex
than the relation between $\nu_0^{\omega}$ and $\langle
\hat{F}_{px} \rangle_{0}^{\omega}$ (see Eq.~\eqref{F25}). Taking
into account the time dependence of the $\langle \hat{F}_{px}
\rangle_{t1}$, it is easy to show that
\begin{equation}
    \label{F44}
    \langle \hat{F}_{px} \rangle_{t1}=\textrm{Re}\{e^{i\omega t}[j^{ac}(Z_1 - 1)]\}.
\end{equation}

Equation~\eqref{F44} gives physical interpretation of the $Z_1$
impedance and its structure may be compared with Eq.~\eqref{F25}.

The real quantities
$\mathbf{E}_1=\textrm{Re}\textbf{\textsl{E}}_1$ and
$\mathbf{j}^{ac}=\textrm{Re}\textbf{\textsl{J}}$ are connected by
the relation $\mathbf{E}_1=\hat{\rho}^{ac}\textbf{j}$, where the
$ac$-response resistivity tensor is
\begin{equation}
    \label{F45}
    \hat{\rho}^{ac}=
        \begin{pmatrix}
            \rho^{ac}_{xx} & \rho^{ac}_{xy} \\
            \rho^{ac}_{yx} & \rho^{ac}_{yy}
        \end{pmatrix}.
\end{equation}
Note that from Eqs.~\eqref{F39} and~\eqref{F42} follows that
\begin{equation}
    \label{F461}
    \hat{\rho}^{ac}=\textrm{Re}\{\hat{Z}(\omega)e^{i\omega t}\}.
\end{equation}


\subsection{Longitudinal and transverse impedance responses}

The experimentally measurable $ac$ resistive responses refer to
coordinate system to the $ac$ current which directed, for
simplicity, at the same angle $\alpha$ with respect to the $y$
axis as the $dc$ current (see Fig.~1). The longitudinal and
transverse (with respect to the $ac$ current direction) components
of the electric field $E^{ac}_{\parallel}$ and $E^{ac}_{\perp}$,
are related to $E_x^{ac}(t)$ and $E_y^{ac}(t)$ by the same
relations as for the $dc$ current (see Eqs.~\eqref{F29}). The
latter is true for the relations between the experimentally
observable longitudinal and transverse (with respect to the
$\mathbf{j}^{ac}$ direction) magnetoresistivities
$\rho^{ac}_{\parallel} = E^{ac}_{\parallel}/j^a$ and
$\rho^{ac}_{\perp} = E^{ac}_{\perp}/j^a$ (where $j^a$ is $j^{ac}$
amplitude ($(j^a)^2=(j^{ac}_x)^2+(j^{ac}_y)^2)$. As a result we
have
\begin{equation}
    \label{F47}
    \rho^{ac}_{\parallel}=\textrm{Re}\{ Z_{\parallel}e^{i\omega t} \},
    \quad \rho^{ac}_{\perp}=\textrm{Re}\{ Z_{\perp}e^{i\omega t} \},
\end{equation}
where
\begin{equation}
    \label{F48}
    \begin{array}{crr}
    Z_{\parallel}=(\rho_f/D)[(D - \delta^2 Z_1)\sin^2\alpha + Z_1\cos^2\alpha],\\
    \\
    Z_{\perp}= (\rho_f/D)[\delta Z_1 - D(1-Z_1)\cos\alpha\sin\alpha].\\
    \end{array}
\end{equation}

Note, however, that the magnitudes of the $\rho^{ac}_{\parallel}$
and $\rho^{ac}_{\perp}$, given by Eqs.~\eqref{F47}, in general (as
in the case of $dc$ current), depend on the direction of the
external magnetic field $\mathbf{B}$ along $z$ axis due to the
$\delta=n\epsilon$ dependence of the $Z_1$ through the implicit
dependence of $\psi_1(\omega)$ on the $j^{ac}$ and $j^{dc}$. In
order to consider only $n$-independent magnitudes of the
$\rho^{ac}_{\parallel}$ and $\rho^{ac}_{\perp}$ resistivities we
should introduce the even (+) and odd (--) longitudinal and
transverse magnetoresistivities with respect to magnetic field
reversal in the form $\rho^{ac\pm}_{\parallel,\perp}(n) \equiv
[\rho^{ac}_{\parallel,\perp}(n) \pm
\rho^{ac}_{\parallel,\perp}(-n)]/2$.

Let us first separate $Z_1=1-i\psi_1/j^{ac}$ on the even
$Z^+_1(n)=Z^+_1(-n)$ and the odd $Z^-_1(n)=-Z^-_1(-n)$ parts. If
we assume $\psi_1(n)=\psi_1^{+}(n)+\psi_1^{-}(n)$, where
$\psi_1^{\pm}(n)$ are the even and odd parts of $\psi_1(n)$ (i.~e.
$\psi_1^{\pm}(n)=[\psi_1(n)\pm\psi_1(-n)]/2$), then we have
\begin{equation}
    \label{F49}
    \begin{array}{crr}
    Z^+_1(n)=1-inj_c[j_y\psi_1^-(n) - \delta j_x\psi_1^+(n)]/(j_y^2-\delta^2 j_x^2),\\
    \\
    Z^-_1(n)=-inj_c[j_y\psi_1^+(n)-\delta j_x\psi_1^-(n)]/(j_y^2-\delta^2 j_x^2).
    \end{array}
\end{equation}

From now on, we can present Eqs.~\eqref{F48} in the form, similar
to Eqs.~\eqref{F31} and ~\eqref{F32}, with the only difference in
the change of $\nu_0^{\omega\pm}$ for $Z_1^{\pm}$ and
$\rho^{dc\pm}_{\parallel,\perp}$ for $Z^{\pm}_{\parallel,\perp}$.
However, hereafter it will be suitable for us to present
Eqs.~\eqref{F48} in another equivalent form
\begin{equation}
    \label{F50}
    Z^+_{\parallel}=(\rho_f/D)[(D - \delta^2Z_1^+)\sin^2\alpha +Z_1^+\cos^2\alpha],
\end{equation}
\begin{equation}
    \label{F51}
    Z^-_{\parallel}=(\rho_f/D)Z_1^-(\cos^2\alpha -\delta^2\sin^2\alpha),
\end{equation}
\begin{equation}
    \label{F52}
    Z^+_{\perp}=(\rho_f/D)[\delta Z_1^- -D(1-Z_1^+)\sin\alpha\cos\alpha],
\end{equation}
\begin{equation}
    \label{F53}
    Z^-_{\perp}=(\rho_f/D)(\delta Z_1^+ + DZ_1^-\sin\alpha\cos\alpha).
\end{equation}


\subsection{The Hall effect and the guiding of vortices in nonlinear AC response}

Let us consider peculiarities of the $ac$ resistive responses in
the investigated model due to the Hall effect. Experimentally,
three types of measurements of the observed $ac$ resistive
characteristics are possible in a prescribed geometry defined by a
fixed value of the angle $\alpha$. First is $ac$ response
measurements which investigate the dependence of observed
$\rho_{\parallel,\perp}^{ac\pm}(\xi^a | \xi^d, g)$ resistivities
on the current density $\xi^a$ at fixed $dc$ current density
$\xi^d$ and temperature $g$. Second is the dependence of
$\rho_{\parallel,\perp}^{ac\pm}(g | \xi^a, \xi^d)$ on the
temperature at fixed $\xi^a$ and $\xi^d$. Third, is the dependence
of $\rho_{\parallel,\perp}^{ac\pm}(\xi^d | \xi^a, g)$ on the $dc$
current density at fixed $\xi^a$ and $g$. The form of these
dependences is governed by a geometrical factor - the angle
$\alpha$ between the directions of the current density vector
$\mathbf{j}(t)$ and the channels of the washboard pinning
potential. There are two different forms of the dependence of
$\rho_{\parallel,\perp}^{ac\pm}$ on the angle $\alpha$ (see
formulas \eqref{F50}-\eqref{F53}). The first of these is the
"tensor" dependence, also present in the linear regimes (similar
to the TAFF and FF regimes for formulas ~\eqref{F31} and
\eqref{F32}), which is external to the impedance $Z_1$ (see
Eqs.~\eqref{F48}). The second is through the dependence of $Z_1$
on its arguments $\xi^a(\alpha)$ and $\xi^d(\alpha)$, which in the
region of the transition from linear in $j^{ac}$ and $j^{dc}$
regimes (at $\xi^{a,d}\ll1$ and $\xi^{a,d}\gg1$) is substantially
nonlinear.

First recall that in the absence of the Hall effect $(\epsilon=0)$
there exist only even (with respect to magnetic field inversion)
impedances $Z^+_{\parallel,\perp}$ -- the odd impedances
$Z^-_{\parallel,\perp}$ are zero (see
Eqs.~\eqref{F50}-\eqref{F53}). The presence of nonzero value of
$\epsilon$ leads not only to the appearance of a Hall contribution
to the observed $ac$ responses on account of the even component
$Z_1^+$ of the impedance $Z_1$, but also to the appearance of the
odd component $Z_1^-$, which has a maximum in the region of the
nonlinear transition from the one linear regime (at low $\xi^a,
\xi^d$ and $g\gg1$) to another linear regime (at large $\xi^a,
\xi^d$ and arbitrary $g$) and is essentially equal to zero outside
this transitional region (see Figs.~10,~11 in Ref.~[3]). As a
consequence, "crossover" effects arise: contributions from $Z_1^-$
to effects due to $Z_1^+$, and vice versa; contributions from
$Z_1^+$ to effects due to $Z_1^-$. Thus, in the even impedance
$Z_{\perp}^+$ (see Eq.~\eqref{F52}), in addition to the main
contribution created by the guiding of vortices and described by
$Z_1^+$ there is present a Hall contribution arising due to
$Z_1^-$. The expression for the odd impedance $Z_{\perp}^-$ (see
Eq.~\eqref{F53}) contains, in addition to the Hall term arising
due to $Z_1^+$, term due to $Z_1^-$.


\subsection{AC response in LT geometries}

In order to study a more simple form of
Eqs.~\eqref{F50}-\eqref{F53} we consider first the $L$ and $T$
geometries of the $ac$ response (see Fig.~1, insert).

In $L$ geometry $\alpha=\pi/2$ and $j_L^{ac} = \epsilon\xi^a$ does
not depend on $n$ as well as $j_L^{dc} = \epsilon\xi^d$ for the
$dc$ current. As a result, $\psi_{1,L}$ as well as
$Z_{1,L}=1-i\psi_{1,L}/j_L^{ac}$ have $\psi_{1,L}^-=Z_{1,L}^-=0$.
Finally, from Eqs.~\eqref{F50} follows
\begin{equation}
    \label{F54}
    Z_{\parallel,L}^+=(\rho_f/D)[1+\delta^2(1-Z_{1,L}^+)], \quad
    Z_{\parallel,L}^-=0,
\end{equation}
\begin{equation}
    \label{F55}
    Z_{\perp,L}^+=0, \quad
    Z_{\perp,L}^-=\delta(\rho_f/D)Z_{1,L}^+.
\end{equation}
If we define $\rho_{\parallel,L}^{ac+}$ and
$\zeta_{\parallel,L}^{ac+}$ as the resistivity and reactivity of
$Z_{\parallel,L}^+$ impedance, respectively, by the relation
$Z_{\parallel,L}^+=\rho_{\parallel,L}^{ac+} -
i\zeta_{\parallel,L}^{ac+}$ we can show that
\begin{equation}
    \label{F56}
    \begin{array}{ccr}
    \rho_{\parallel,L}^{ac+}=(\rho_f/D)(1-\epsilon\textrm{Im}\psi_{1,L}/\xi^a), \\
    \\
    \zeta_{\parallel,L}^{ac+}=-(\rho_f/D)\epsilon\textrm{Re}\psi_{1,L}/\xi^a.\\
    \end{array}
\end{equation}

Note that experimentally measured quantities $|Z_{\parallel,L}^+|$
and
$\tan\varphi_{\parallel,L}=\zeta_{\parallel,L}^{ac+}/\rho_{\parallel,L}^{ac+}$
allow to obtain $\rho_{\parallel,L}^{ac+}$ and
$\zeta_{\parallel,L}^{ac+}$ and to compare them with the
theoretical formulas ~\eqref{F56}. Similar calculations for
$Z_{\perp,L}^-=\rho_{\perp,L}^{ac-} -i\zeta_{\perp,L}^{ac-}$ yield
\begin{equation}
    \label{F57}
    \begin{array}{ccr}
    \rho_{\perp,L}^{ac-}=n(\rho_f/D)(\epsilon\xi^a-\textrm{Im}\psi_{1,L})/\xi^a,\\
    \\
    \zeta_{\perp,L}^{ac-}=n(\rho_f/D)\textrm{Re}\psi_{1,L}/\xi^a.\\
    \end{array}
\end{equation}

In $T$ geometry (see insert in Fig.~1) $\alpha=0$,
$j_T^{ac}=n\xi^a$ and $j_T^{dc}=n\xi^d$. In this case it follows
from Eqs.~\eqref{F17}-\eqref{F20} that
$\psi_{1,T}(n)=-\psi_{1,T}(-n)$, i.~e. $\psi_{1,T}$ is an odd
function of $n$ and $\psi_{1,T}^+=0$. As a result,
\begin{equation}
    \label{F58}
    Z_{1,T}^+=1-in\psi_{1,T}^-/\xi^a, \quad Z_{1,T}^-=0.
\end{equation}
Then from Eqs.~\eqref{F50}-~\eqref{F53} we have
\begin{equation}
    \label{F59}
    Z_{\parallel,T}^+=(\rho_f/D)Z_{1,T}^+, \quad
    Z_{\parallel,T}^-=0,
\end{equation}
\begin{equation}
    \label{F60}
    Z_{\perp,T}^-=(\rho_f/D)\delta Z_{1,T}^+, \quad
    Z_{\perp,T}^+=0.\\
\end{equation}

If $Z_{\parallel,T}^+=\rho_{\parallel,T}^{ac+} -
i\zeta_{\parallel,T}^{ac+}$ and
$Z_{\perp,T}^-=\rho_{\perp,T}^{ac-} - i\zeta_{\perp,T}^{ac-}$,
then from Eqs.~\eqref{F59},~\eqref{F60} we obtain
$Z_{\perp,T}^-=\delta Z_{\parallel,T}^+$ and
\begin{equation}
    \label{F61}
    \rho_{\perp,T}^{ac-}=\delta \rho_{\parallel,T}^{ac+}, \quad
    \zeta_{\perp,T}^{ac-}=\delta \zeta_{\parallel,T}^{ac+}.
\end{equation}

From Eqs.~\eqref{F61} follows that that experimentally measured
quantities satisfy the simple relations
\begin{equation}
    \label{F62}
    |Z_{\perp,T}^-|=\epsilon |Z_{\parallel,T}^+|, \quad
    \tan\varphi_{\perp,T}^-=\tan\varphi_{\parallel,T}^+.
\end{equation}

At last, from Eqs.~\eqref{F58} and \eqref{F59} follows that
$\rho_{\parallel,T}^{ac+}=(\rho_f/D)\rho_{1,T}^{ac+}$ and
$\zeta_{\parallel,T}^{ac+}=(\rho_f/D)\zeta_{1,T}^{ac+}$, where
\begin{equation}
    \label{F63}
    \begin{array}{ccr}
    \rho_{1,T}^{ac+}=1+n\rm{Im}\psi_{1,T}^-/\xi^a,\\
    \\
    \zeta_{1,T}^{ac+}=n\rm{Re}\psi_{1,T}^-/\xi^a, \quad \rho_{1,T}^-=\zeta_{1,T}^-=0.\\
    \end{array}
\end{equation}


\subsection{The power absorption in AC response}

In order to calculate the power absorbed per unit volume
$\mathcal{\bar{P}}$ (and averaged over the period of an $ac$
cycle) we use the standard relation
$\mathcal{\bar{P}}=(1/2)\textrm{Re}(\textbf{\textsl{E}}_1\cdot\textbf{\textsl{J}})$
where $\textbf{\textsl{E}}_1$ and $\textbf{\textsl{J}}$ are the
complex amplitudes of the $ac$ electric field and current density,
respectively. Using Eqs.~\eqref{F43} and~\eqref{F48} we can show
that
\begin{equation}
    \label{F64}
    \mathcal{\bar{P}}=(j^2/2)\bar{\rho}\equiv(j^2/2)\textrm{Re}Z_{\parallel}.
\end{equation}
After some algebra we obtain that
\begin{equation}
    \label{F65}
    \bar{\rho}=(\rho_f/D)[D\sin^2\alpha + (1-D\sin^2\alpha)\textrm{Re}Z_1.
\end{equation}

Taking into account that $Z_1\equiv1 -iG_1$, where
$G_1\equiv\psi_1/j^{ac}$, we conclude that
$\textrm{Re}Z_1=1-\textrm{Re}(iG_1)=1 + \textrm{Im}G_1$. Then from
Eq.~\eqref{F65} we have
\begin{equation}
    \label{F66}
    \bar{\rho}=(\rho_f/D)[1 + (1-D\sin^2\alpha)\textrm{Im}G_1].
\end{equation}
In the limit of $\epsilon\ll1$ we obtain for $\bar{\rho}$ a more
simple result
\begin{equation}
    \label{F67}
    \bar{\rho}=\rho_f(1 + \textrm{Im}G_1\cos^2\alpha),
\end{equation}
which will be analyzed in detail in Sec.~IV.~F.

From Eq.~\eqref{F65} follows two simple results for $\bar{\rho}$
in $LT$ geometries
\begin{equation}
    \label{F68}
    \bar{\rho}_L=(\rho_f/D)(D-\delta^2\textrm{Re}Z_{1L})=(\rho_f/D)(1-\epsilon\textrm{Im}\psi_{1L}/\xi^a),
\end{equation}
\begin{equation}
    \label{F69}
    \bar{\rho}_T=(\rho_f/D)\textrm{Re}Z_{1T}=(\rho_f/D)(1+n\textrm{Im}\psi_{1T}^-(n)/\xi^a).
\end{equation}
Note that $\bar{\rho}_L$ and $\bar{\rho}_T$ in Eqs.~\eqref{F68}
and~\eqref{F69} are equal to the expressions for the
$\rho^{ac+}_{\parallel,L}$ and $\rho^{ac+}_{\parallel,T}$ given by
Eqs.~\eqref{F56} and ~\eqref{F63}, respectively.


\subsection{AC impedance and power absorption at $\epsilon\ll1$}

Here we analyze the $ac+dc$ impedance dependences
$Z^{\pm}_{\parallel}$ and $Z^{\pm}_{\perp}$ (see
Eqs.~\eqref{F50}-\eqref{F53}), with allowance for the small Hall
effect $(\epsilon\ll1)$. In this case Eqs.~\eqref{F50}-\eqref{F53}
become more simple and the expressions for
$Z^{\pm}_{\parallel,\perp}$, out of terms of order
$\epsilon^2\ll1$, have the form
\begin{equation}
\label{F73}
    \begin{array}{crr}
    Z^+_{\parallel}=\rho_f(Z_1^+\cos^2\alpha+\sin^2\alpha),\\
    \\
    Z^+_{\perp}=\rho_f(Z_1^+ -1)\sin\alpha\cos\alpha,
    \end{array}
\end{equation}
\begin{equation}
\label{F74}
    \begin{array}{crr}
    Z^-_{\parallel}=\rho_fZ^-\cos^2\alpha,\\
    \\
    Z^-_{\perp}=\rho_f(\delta Z_1^+ +Z_1^-\sin\alpha\cos\alpha),
    \end{array}
\end{equation}

Here $Z_1^{\pm}$ can be obtained from relations
\begin{equation}
    \label{F75}
    \begin{array}{ccr}
    Z_1=Z_1^+ + Z_1^- = 1-iG_1=\\
    \\
    1-i(G_1^+ + G_1^- )=(1-iG_1^+) - iG_1^-,
    \end{array}
\end{equation}
where $G_1\equiv\psi_1(j^{ac}, j^{dc})/j^{ac}$ and
$Z_1^+=(1-iG_1^+), Z_1^-=-iG_1^-$. In the case of a small Hall
effect $(\epsilon\ll1)$ the expression for even (+) and odd (--)
components of $Z_1(\xi^{ac}, \xi^{dc})$ (in terms of $G_1^{\pm}$)
in the linear approximation in the parameter
$\epsilon\tan\alpha\ll1$ are equal, respectively, to
\begin{equation}
\label{F76}
    \begin{array}{crr}
    G_1^+=G_1(n\xi_y^a, n\xi_y^d), + nR_a^+\delta\tan\alpha\\
    \\
    G_1^-=(- G_1^+ + nR_a^- )\delta\tan\alpha,
    \end{array}
\end{equation}
\begin{equation}
\label{F77}
    R_a\equiv[(\partial\psi_1/\partial\xi_y^a) + (j^d/j^a)(\partial\psi_1/\partial\xi_y^d)],
\end{equation}
where $\psi_1=\psi_1(n\xi_y^a, n\xi_d^a)$, $j^a$ and $j^d$ are
$ac$ and $dc$ current density values, and $R_a^+$, $R_a^-$ are
even and odd parts of the $R_a$, respectively.

It is worth noticing that Eqs.~\eqref{F73} and ~\eqref{F74} have
the same structure as Eqs.~\eqref{G4} and~\eqref{G5} for the
$dc+ac$ response at $\epsilon\ll1$. Actually, if we change
$Z_1^{\pm}$ and $Z^{\pm}_{\parallel,\perp}$ in Eqs.~\eqref{F73}
and~\eqref{F74} by $\nu_0^{\omega\pm}$ and
$\rho^{dc\pm}_{\parallel,\perp}$, respectively, we obtain then
Eqs.~\eqref{G4} and~\eqref{G5}.

So all conclusions following the discussion about a structure of
these equations can be repeated for the Eqs.~\eqref{F73}
and~\eqref{F74}.

It is interesting also to analyze an anisotropic power absorption
in the limit of $\epsilon\ll1$, given by Eq.~\eqref{F67} in the
previous Section IV.~E.

Let us put $G_1=G_1^+ + G_1^-$, where $G_1^{\pm}$ are presented by
Eqs.~\eqref{F76} and~\eqref{F77}. In case where $\epsilon=0$,
$G_1=G_1^+=G_1(n\xi_y^a,n\xi_y^d)=\psi_1(n\xi_y^a,n\xi_y^d)/n\xi_y^a$,
where $\xi_y^a=\xi^a\cos\alpha$, $\xi_y^d=\xi^d\cos\alpha$ and

\begin{equation}
    \label{F78}
    \bar{\rho}=\rho_f[1 + \cos\alpha\cdot n\textrm{Im}\psi_1^-(n)/\xi^a],
\end{equation}

Note, that Eq.~\eqref{F78} at $\alpha=0$ yields
$\bar{\rho}=\bar{\rho}_T(\epsilon=0)$, where $\bar{\rho}_T$ is
given by Eq.~\eqref{F69}.


\subsection{Linear ac response}

Here we assume that $j=j^{dc} + j^{ac}e^{i\omega t}$ and the
alternating current is small ($j^{ac}\ll1$). There are
\textit{three} different ways to derive \textit{linear} (in
$j^{ac}$) impedance $Z_{1l}$ at arbitrary value of $j^{dc}$.

The first way is to use general expression for $Z_1(j^{ac},
j^{dc})$ (see Eq.~\eqref{F38}) derived by the method of
\textit{matrix} continued fraction at arbitrary magnitudes of the
$j^{ac}$ and $j^{dc}$. If we take into account that
$\psi_1(j^{ac}=0)=0$, then it follows
\begin{equation}
    \label{L57}
    Z_{1L} = \lim_{\substack{j^{ac}\rightarrow0}}Z_1 = 1 - i
    (d\psi_1/dj^{ac})|_{j^{ac}=0}.
\end{equation}
This method is the most general and powerful if we can calculate
$Z_1(j^{ac},j^{dc})$.

The second way is to calculate $Z_{1l}$ by means of making the
perturbation expansion of the $\langle r^m\rangle(t)$ (see
Eq.~\eqref{F7}) in powers of $j^{ac}\ll1$ in the form
\begin{equation}
    \label{L58}
    \langle r^m\rangle_t = \langle r^m\rangle_0 + \langle
    r^m\rangle_1 + \dots,
\end{equation}
where $\langle r^m\rangle_1=A_m(\omega)j^{ac}e^{i\omega t}$ and
the subscript $"0"$ denotes the statistical averages in the
absence of the $ac$ and the subscription $"1"$ the portion of the
statistical average which is linear in the $ac$. Whereas the
$\langle r^m\rangle_0$ satisfies Eq.~\eqref{D29}, the complex
amplitude $A_m(\omega)$ (for $m\geq1$) can be presented (see
details in subsection 5.5 of Ref.~[\cite{CKW}]) in terms of the
infinite scalar continued fraction $\tilde{S}_m(\omega)$ as
\begin{equation}
    \label{L59}
    A_1(\omega) = 2i
    \sum_{\substack{n=1}}^{\substack{\infty}}(-1)^n\prod_{\substack{m=1}}^{\substack{n}}S_m\tilde{S}_m(\omega),
\end{equation}
where
\begin{equation}
    \label{L60}
    \mathbf{\tilde{S}}_m(\omega)=\cfrac{1/2}{\frac{i\omega\hat{\tau}}{m} + ij^{dc} + \frac{m}{g} +\cfrac{1/4}
    {\frac{i\omega\hat{\tau}}{m+1} + ij^{dc} + \frac{m+1}{g} +...}}
\end{equation}

and $\tilde{S}_m(\omega=0) = S_m$ (see also Eqs.~\eqref{D30} and
\eqref{D31}). Using Eq.~\eqref{L59} and taking into account that
$A_{-1}(\omega)=-A^{\ast}_1(-\omega)$ we conclude that
\begin{equation}
    \label{L61}
    \langle\sin\textsl{x}\rangle_1 = (i/2)(\langle r\rangle_1 - \langle
    r^{-1}\rangle_1) = B(\omega)j^{ac}e^{i\omega t},
\end{equation}
where $B(\omega)\equiv(i/2)[A_1(\omega) + A^{\ast}_1(-\omega)]$.
Then from expressions for the $\langle v_x\rangle_{t1}$, taken at
$j^{ac}e^{i\omega t}$, Eqs.~\eqref{F44} and \eqref{L61} follows
that dimensionless linear impedance is
\begin{equation}
    \label{L62}
    Z_{1l}(\omega) = 1 - B(\omega).
\end{equation}
At last, the third way to calculate the linear impedance gives an
approximate \textit{analytical} expression for $Z_{1l}(\omega)$
within the frames of the method of effective eigenvalue (see
details in subsections 5.6 and 5.7 of Ref.~[\cite{CKW}]).
Following this approach we can express the dimensionless linear
impedance in terms of the modified Bessel functions $I_{\nu}(z)$
as
\begin{equation}
    \label{L63}
    Z_{1l}(\omega,g,j^{dc}) = 1 -
    \frac{1}{2}[\frac{I_{1+\mu}(g)}{I_{\mu}(g)
    (\lambda + i\omega\hat{\tau})} + \frac{I_{1-\mu}(g)}{I_{-\mu}(g)(\lambda^{\ast} +
    i\omega\hat{\tau})}],
\end{equation}
where
\begin{equation}
    \label{L64}
    \lambda = I_{\mu}(g)I_{1+\mu}(g)/[2\int_{\substack{0}}^{\substack{g}}I_{\mu}(t)I_{1+\mu}(t)dt]
\end{equation}
is an effective eigenvalue~\cite{CKW} and $\mu \equiv igj^{dc}$.
It follows from Eqs.~\eqref{L63} and ~\eqref{L64} that at
$\omega=0$
\begin{equation}
    \begin{array}{ccr}
    \label{L65}
    Z_{1l}(\omega=0,g,j^{dc}) = d[j^{dc}\nu_0(j^{dc})]/dj^{dc} =\\
    \\
    =1-\textrm{Re}[(2/I_{\mu}^2(g))\int_{\substack{0}}^{\substack{g}}I_{\mu}(t)I_{1+\mu}(t)dt],\\
    \end{array}
\end{equation}
where $\nu_0(j^{dc})$ is given by Eq.~\eqref{D32}. Note also that
the right-hand side of Eq.~\eqref{L65} is the exact expression for
the dimensionless static differential resistivity in an analytical
form. In the limit $j^{dc}=0$ from Eq.~\eqref{L65} follows the
well-known result of Coffey and Clem$^{11}$ (see also
Refs.~[\cite{MART, IZR}]). Actually, in this limit $\mu=0$ and
\begin{equation}
    \label{L66}
    \lambda = \lambda^{\ast} = I_0(g)I_1(g)/[I_0^2(g)-1].
\end{equation}
As a result
\begin{equation}
    \label{L67}
    Z_{1l}(\omega,g,j^{dc}=0)\equiv Z_{1l}^0=\frac{\nu_{00}+(\omega\tau)^2
    + i\omega\tau(1-\nu_{00})}{1+(\omega\tau)^2},
\end{equation}
where $\nu_{00}\equiv\nu_0(j^{dc}=0)=1/I_0^2(g)$ is the flux creep
factor\cite{CLEM,MART,SD,SSS} and
\begin{equation}
    \label{L68}
    \tau=\hat{\tau}/\lambda=\hat{\tau}[I_0^2(g)-1]/I_0(g)I_1(g) =
    \hat{\tau}(1-\nu_{00})[I_0(g)/I_1(g)]
\end{equation}
is the characteristic relaxation time. If $Z_{1l}^0=\rho_{1l}^0 -
i\zeta_{1l}^0$, where $\rho_{1l}^0$ and $\zeta_{1l}^0$ are linear
resistivity and reactivity in the absence of the $dc$ current,
respectively, then from Eq.~\eqref{L67} (see also Eq.~\eqref{I})
follows that
\begin{equation}
    \label{L69}
    \rho_{1l}^0(\omega,g)=1-\frac{1-\nu_{00}}{1+(\omega\tau)^2},
    \qquad \zeta_{1l}^0 = -
    \frac{\omega\tau(1-\nu_{00})}{1+(\omega\tau)^2}.
\end{equation}
As expected, in the limit of zero temperature
($g\rightarrow\infty$) we have that $\nu_{00}\rightarrow0,
\tau\rightarrow\hat{\tau}$ and the results of Gittlemann and
Rosenblum\cite{GR} (see also Eqs.~\eqref{I1} and \eqref{I}) are
following from Eqs.~\eqref{L69}.


\subsection{Nonlinear impedance and harmonics response}

Let us consider strong nonlinear effects in the $ac$ impedance of
a sample subjected to a pure $ac$ drive dimensionless current
density $\xi^a\cos\omega t$, where
$\xi^a\equiv|\mathbf{j}^{ac}|/j_c$. In the following we will
discuss the behavior of $(\xi^d,~\xi^a,~\Omega,~g)$-dependent
impedance for simplicity in terms of the dimensionless $ac$
resistivity $\rho_1^{ac+}=\rho_1$ and reactivity
$\zeta_1^{ac+}=\zeta_1$. As the angular $\alpha$-anisotropy in
these responses is omitted, the experimental observation of the
following dependences (see Figs. 4-8) can be carried out in fact
by the measurement of the $\rho_{\parallel,T}^{ac+}$ and
$\zeta_{\parallel,T}^{ac+}$ responses in \emph{T} geometry (see
Eqs.~(77),~(78), and the definition of the $Z_{\parallel,T}^+$).
Figure~4 shows the dimensionless $ac$ resistivity $\rho_1$ and
reactivity $\zeta_1$ versus $ac$ current density $\xi^a$ for
different dimensionless frequencies $\Omega\equiv\omega\hat{\tau}$
at very low temperature (g=100).

As can be seen from the Fig.~4(a), when $\Omega$ is very small,
the $\rho_1(\xi^a)$ shows several characteristic features: a
threshold $\xi^a_c$ value and a subsequent parabolic rise, above
the threshold, with associated steplike structures. The threshold
current density where a sudden increase in $\rho_1(\xi^a)$ starts
may be defined as critical current density $\xi_c^a$. The step
height decreases with $\xi^a$ increasing. The reactivity
$\zeta_1(\xi^a)$ shows nearly periodic dynamic $2\pi$ -jumps of
the vortex coordinate occurring as the drive current density
$\xi^a$ is increased (see Fig.~4(b)). The curves in Figs.~4(a,~b)
look like the similar curves discussed earlier~\cite{ZHAI} for the
nonlinear resistance and reactance of the purely $ac$-driven
resistively shunted Josephson junction model at $T=0$ where the
overall shape and phase slips of these curves at several
dimensionless frequencies was explained in terms of the
bifurcations in the time-dependent solution of the equation for
the phase difference $\varphi$ across the junction~\cite{MDC,
WOS}. Analogous bifurcations of the dimensionless coordinate
$\textsl{x}$ versus dimensionless time $\hat{\tau/}\pi$ in our
problem at $T=0$ can be calculated too.

\begin{figure}[t]
    \epsfig{file=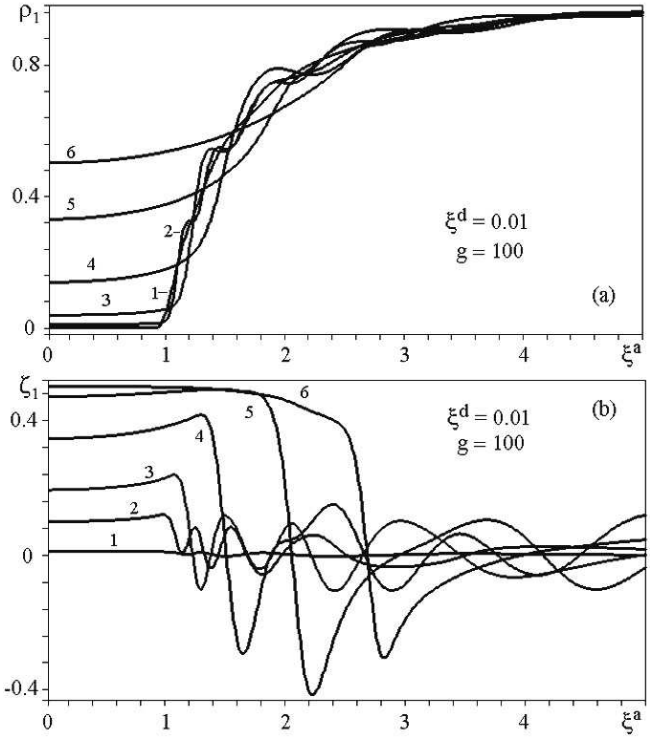,width=0.48\textwidth}
    \caption{The $ac$ resistivity $\rho_1$ and reactivity $\zeta_1$ versus $\xi^a$ for various $\Omega=0.01(1), 0.1 (2), 0.2 (3), 0.4 (4), 0.7 (5), 1 (6), \xi^d=0.01, g=100.$} \label{fig2}
\end{figure}

These bifurcations can cause sudden changes in $E_{y1}^{ac}(t)$
during one cycle of the alternating current and hence result in
steps~\cite{ZHAI}.When $\Omega$ becomes large, both the threshold
and steps in $\rho_1(\xi^a)$ disappear and the amplitude of the
$\textsl{x}$-jump in $\zeta_1(\xi^a)$ becomes larger. Also, the
$\textsl{x}$-jump moves to large values of $\xi^a$ and the spacing
in $\xi^a$ between bifurcations becomes large which results in
$\rho_1$ and $\zeta_1$ approaching unity. Because in our problem
the abrupt $2\pi$-jumps of the dimensionless vortex coordinate
$\textsl{x}$ correspond to the overcoming by vortex of the
potential barrier between two neighboring potential wells at
nonzero temperature, our curves $\rho_1(\xi^a)$ and
$\zeta_1(\xi^a)$, in comparison with the curves of
Refs.~[\cite{MDC, WOS}] are smoothed due to the influence of a
finite temperature.

It is worth noticing that the magnitude of $\rho_1(\xi^a|\Omega)$
in Fig.~4(a) at $\xi^a<1$ is approximately equal to a constant
which progressively increases with $\Omega$ increasing. From a
physical viewpoint it corresponds to the enhancement of power
absorption with the growth of $\Omega$ due to the increasing of
the viscous losses accordingly to GR (see Introduction) mechanism.

Now we consider the case when an $ac$ current driven sample is
$dc$ current biased, i.~e. the washboard pinning potential is
tilted. In Fig.~5 we plot $\rho_1$ and $\zeta_1$ versus $\xi^a$
for various values of $\xi^d\equiv|\mathbf{j}^{dc}|/j_c$ at fixed
dimensionless frequency $\Omega=0.1$ and very low temperatures
(g=100). There are two regimes of behavior noticeable,
corresponding to $\xi^d>\xi^a$ or $\xi^a<\xi^d$.

\begin{figure}[t]
    \epsfig{file=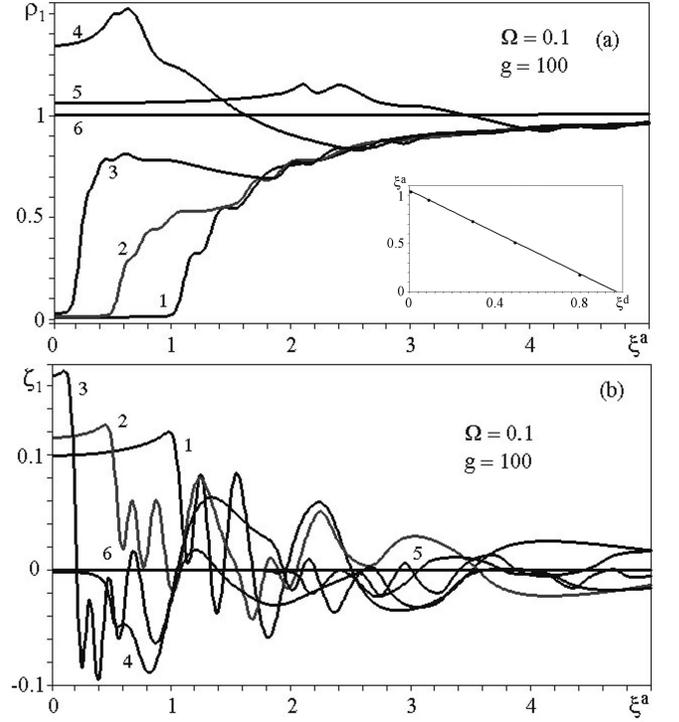,width=0.48\textwidth}
    \caption{The $ac$ resistivity $\rho_1$ and reactivity $\zeta_1$ versus $\xi^a$ for various $\xi^d=0.01 (1), 0.5 (2), 0.8(3), 1.5 (4), 3 (5), 10 (6), g=100, \Omega=0.1$.} \label{fig2}
\end{figure}
\begin{figure}[t]
    \epsfig{file=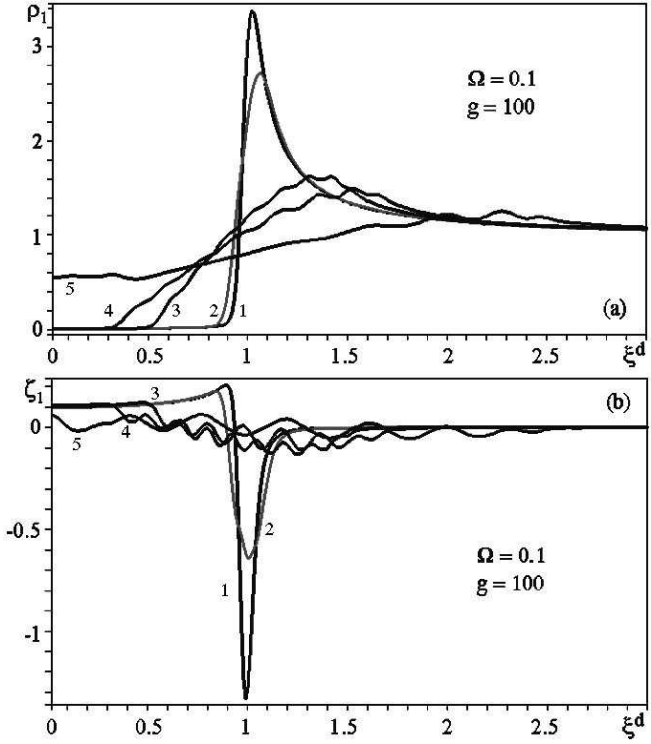,width=0.48\textwidth}
    \caption{The $ac$ resistivity $\rho_1$ and reactivity $\zeta_1$ versus $\xi^d$ for various $\xi^a=0.01 (1), 0.1 (2), 0.5 (3), 0.7 (4), 1.5 (5), 10 (6), g=100, \Omega=0.1$.} \label{fig2}
\end{figure}

When $\xi^d$ is very small (for instance, $\xi^d=0.01$), the
critical $ac$ current density $\xi_c^a(\xi^d=0.01)$ is found to be
equal to $A\approx1$(see Fig.~5(a)). When the $dc$ drive current
density $\xi^d$ is smaller than $A$ (i.~e. $0<\xi^d<A$), it can be
seen that the $ac$ critical current density $\xi_c^a$ as a
function of $\xi^d$ decreases and that both the step size and step
rising pattern are changed. In the inset to Fig.~5(a) we plot the
$\xi^a_c$ as a function of $\xi^d$ for $\Omega=0.1$. A linear fit
$\xi_c^a=-\xi^d + A$ yields $A=1.02$. Note however that this
$\xi_c^a(\xi^d)$ function is weakly frequency-dependent.

When $\xi^d>A$, initially apparent jumps appear and with further
increase of the $ac$ drive current density $\xi^a$, the $\rho_1$
begins to decrease with smoothed intermittence steps and
eventually approaches the unity. This regime
$(\xi^a,\xi^d)>\xi_c^a$ is rather interesting because it shows
strong vortex-locking effects in the $ac$ impedance, similar to
the Shapiro steps seen in the $dc$ CVC`s. For very large values of
$\xi^d$ ($\xi^d=10$), as expected, the effect of the microwave
current density is negligible and the $ac$ dimensionless
resistivity $\rho_1$ approaches the unity.

In the case of $ac$ reactivity $\zeta_1$, which is plotted in
Fig.~5(b), the smoothed $\textsl{x}$-jump is not affected by the
increase in the $dc$ current density $\xi^d$, however, the
amplitude of this jump reduces substantially for larger values of
$\xi^d$.

In Fig.~6 we plot $\rho_1$ and $\zeta_1$ versus $\xi^d$ for
different $ac$ drive current densities $\xi^a=0.01, 0.1, 0.5, 0.7,
1.5, 10$. It can be observed from the Fig.~6(a) that for all
values of $\xi^a$, as $\xi^d$ increases, $\rho_1$ initially
increases, reaches a maximum (very narrow and high for
$\xi^a\lesssim0.1$) and approaches unity eventually. On the other
hand, $\zeta_1$, decreases for $\xi^a\lesssim0.1$ slowly increases
below $\xi^d\lesssim1$ and then sharply decreases, having in the
vicinity of $\xi^d\approx1$ deep minimum, and then approaches to
zero. So the occurrence of the $\textsl{x}$-jump can be seen
clearly when $\xi^d$ is small, whereas large values of either
$\xi^a$ or $\xi^d$ diminish the effect of the $\textsl{x}$-jumps.

\begin{figure}[t]
    \epsfig{file=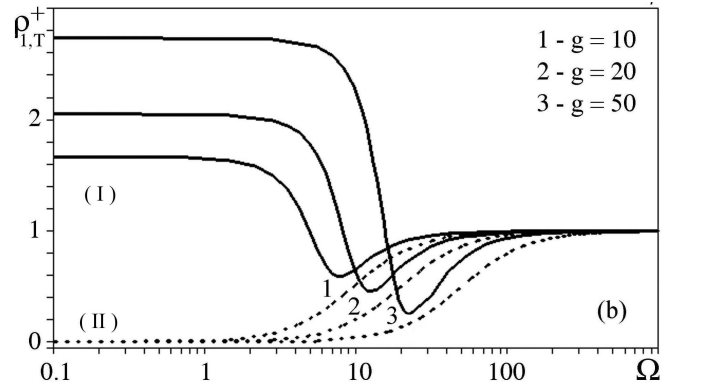,width=0.48\textwidth}
    \caption{ The frequency dependence of $\rho^+_{1,\textrm{T}}(\Omega)$ (b) for various $g$, (I)~$\xi^d=1$~(solid lines), (II) $\xi^d=0$~(dotted lines).} \label{fig2}
\end{figure}

In Fig.~6(a) the $\rho_1(\xi^d)$ dependences demonstrate several
main features. First, the curves, calculated at $\xi^a\ll1$ show
the progressive shrinking of the flux creep range (where the
$\rho_1(\xi^d)\lll1$) with the $\xi^a$ increasing. If we define
the $\xi^d_c(\xi^a)$ as the dependence of the $dc$ critical
current on the value of a small $ac$ driving, then we can show
that $\xi^d_c\simeq1-\xi^a$ at $\xi^a\ll1$. The physical reason
for such behavior is obvious. Second, the appearance of a high
peak in $\rho_1(\xi^d)$ near $\xi^d=1$ for $\xi^a\rightarrow0$ can
be simply explained from an examination of the $dc$ CVC curves,
calculated at $\xi^a\rightarrow0$. In this limit it is evident
that a \emph{dynamic} $dc$ resistivity (taken in the vicinity of
the $\xi^d_c$), which equals to the derivative of the $dc$ CVC
with respect to the $\xi^d$, is strongly enhanced at
$T\rightarrow0$. Third, taking into account an analogy between
Brownian motion in a tilted periodic potential and continuous
phase transitions~\cite{USA}, one can say that a threshold type
phase transition in the vortex motion along the $x$-axis occurs
between the "localized" vortex state at $\xi^d<\xi^d_c(\xi^a)$ and
the "delocalized" running state at $\xi^d>\xi^d_c(\xi^a)$. If we
consider only the \textit{linear} impedance response (i.~e.
$Z_{1,L}(\omega)$ does not depend on $\xi^a$), this phase
transition takes place at $\xi^d=1$ and at the
$x$-point~\cite{USA} where $d^2U_p/dx^2=0$. Forth, as it was shown
recently in~\cite{RIEN}, a strong enhancement of the effective
diffusion coefficient D of an overdamped Brownian particle in a
tilted washboard potential near the critical tilt may occur; that,
in our case, $\textrm{D}(\xi^d)$ may have a peak in the vicinity
of $\xi^d=1$.

The consequences of the D-enhancement we analyze with the aid of
Fig.~7 where the frequency dependence of $\rho_1(\Omega|\xi^d=0)$
(monotonic curves) and $\rho_1(\Omega|\xi^d=1)$ (nonmonotonic
curves) calculated at $\xi^a=0$ for three different temperatures
($g=10, 20, 50$) are shown. The monotonic curves $\xi^d=0$ agree
with the results of Coffey and Clem~\cite{CLEM} who, in fact,
calculated the temperature dependence of the depinning frequency
(introduced at $T=0$ in~\cite{GR}) in a \emph{nontilted }cosine
pinning potential. In contrast to this monotonic behaviour, the
nonmonotonic curves ($\xi^d$=1) demonstrate two characteristic
features. First, an anomalous power absorption
($\rho_1\simeq1,6\div2,8$) at very low frequencies. Second, a deep
minimum for the power absorption ($\rho_1\simeq0,3\div0,6$) at
$T$-dependent $\Omega_{min}$. The appearance of this frequency-
and temperature dependent minimum at $\xi^d=1$ may be related to
the resonance activated reduction of the mean escape time of the
Brownian particle due to an oscillatory variation of the pinning
barrier height~\cite{BOG}.

At last we consider the frequency dependence of the k-th
transformation coefficient amplitude, i.e. $|Z_k|(\Omega | j^d,
j^a,g )$ taken at different values of the \emph{dc} and \emph{ac}
current densities and inverse temperature. Here we point out only
the summary of the $|Z_k|(\Omega)$ curves behaviour because a more
detailed description of these results (interesting for
applications) will be published elsewhere.

\begin{figure}[t]
   \epsfig{file=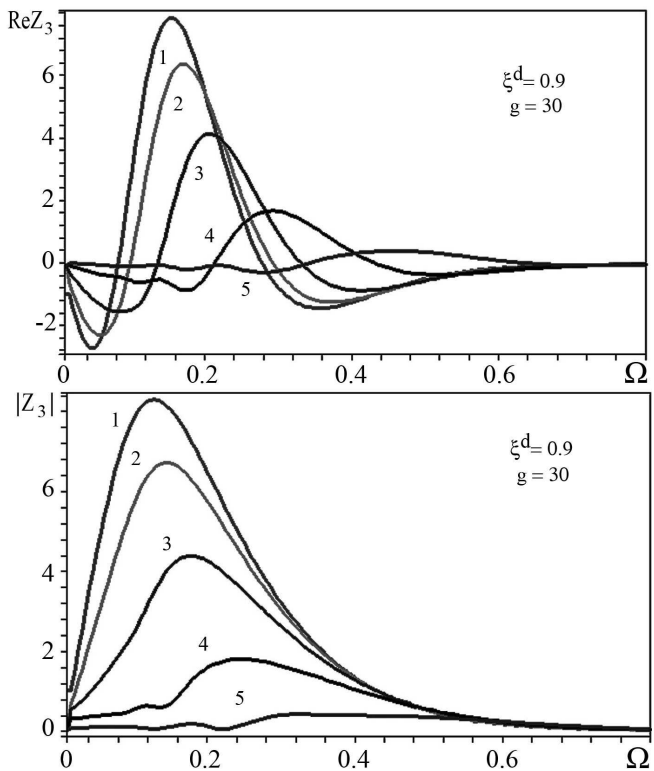,width=0.48\textwidth}
    \caption{The third harmonic response $\textrm{Re}Z_3$ and $|Z_3|$
    versus $\Omega$ for various $\xi^a=0.01(1), 0.1 (2), 0.3 (3), 0.5 (4), 0.7 (5), \xi^d=0.9, g=30.$}\label{fig2}
\end{figure}

The main feature of the frequency dependence of the k-th harmonics
is the appearance (see Fig.~8) of a pronounced maximum at
$\xi^d=1$ for $\Omega_{max}=0.1$  in both
$\textrm{Re}{Z_k(\Omega)}$ and $|Z_k|(\Omega)$ curves. The
magnitude of the maximum is increasing with $\xi^a$-decreasing to
zero and $g$-increasing to infinity. For example, the magnitudes
of $\textrm{Re}{Z_k(\Omega_{max})}$ and $|Z_k|(\Omega_{max})$ are
approximately equal to 20 for $k=3,4$ at $g=30$, $\xi^d=1$ and
$\xi^a=0.01$.The emergence of this maximum and its growth with the
temperature decreasing ($g\rightarrow\infty$) is related to the
origin of a singularity of the $\delta'(\omega)$-type in the
$\Omega$-dependence of the linear impedance response of the
overdamped shunted Josephson junction at T=0 and
$\xi^d\lesssim1$~\cite{AUR}. Note also that
$\textrm{Re}{Z_k(\Omega)}$ becomes negative in the vicinity of
$\Omega_{max}$ which, in turn, is related to the origin of a
similar singularity in the linear $Z_{1}(j^{dc}| { \Omega},g)$
response with the $\Omega$-increasing.


\section{Conclusion}

In conclusion, the considered exactly solvable two-dimensional
model of the vortex  dynamics is of great interest since a very
rich physics is expected from combination of a  strong $dc$ and
$ac$ driving, arbitrary value of the  Hall effect (note, that a
big Hall effect was observed in YBCO~\cite{HAR}), and the low
temperature mediated vortex hopping (or running) in a washboard
pinning potential. The obtained findings substantially generalize
previous theoretical results in the field of the \emph{dc} [2,3]
and \emph{ac} [11-13] stochastic approach to the study of the
vortex dynamics in the washboard planar pinning potential.
Experimental realization of this model in thin-film
geometry~\cite{HUTH,OKS} opens up a possibility for a variety of
experimental studies of directed motion of vortices under
($dc+ac$)- driving simply by measuring longitudinal and transverse
voltages. Experimental control  of a frequency and value of the
driving forces, damping, Hall constant, pinning parameters and
temperature can be effectively provided.

While the discussion in this paper has been entirely in the
context of nonlinear 2D pinning-mediated vortex dynamics, we are
aware that obtained results are generic to all systems with a
tilted washboard potential subjected to an \emph{ac} driving. In
this sense we are conscious of that physical explanation of our
results should be  supplemented by several new notions widely
discussed. Here we mean notions of stochastic
resonance~\cite{BOG}, resonance activation~\cite{GAM}, noise
enhanced stability~\cite{MAN} which may be used not only for
interpretation of our theoretical results, but on the contrary,
the experimental verification of some predictions of these new
approaches may be performed with the aid of the model under
discussion.

It was shown also how pronounced nonlinear effects appear in the
\emph{ac} response and the linear response solutions are recovered
from the nonlinear \emph{ac} response in the weak \emph{ac}
current limit. An influence of a subcritical or overcritical
\emph{dc} current on the time-dependent stationary \emph{ac}
longitudinal and transverse resistive vortex response (on the
frequency of an \emph{ac}-driving $\Omega$) in terms of the
nonlinear impedance tensor $\hat{Z}$  and a nonlinear \emph{ac}
response at $\Omega$-harmonics are studied. New analytical
formulas for $2D$ temperature-dependent \textit{linear} impedance
tensor $\hat{Z}_L$ in the presence of a $\emph{dc}$ current which
depend on the angle $\alpha$ between the current density vector
and the guiding direction of the washboard PPP are derived and
analyzed. Influence of $\alpha$-anisotropy and the Hall effect on
the nonlinear power absorption by vortices is pointed out.

Up to now we have considered only the vortex motion problem. For
the future experimental verification of our theoretical findings
we should keep in mind that they may be applied directly only for
thin-film superconductors in the form of naturally grown (for
example, in the untwined \emph{a}-axis oriented YBCO
film~\cite{TRA}) and artificially prepared washboard pinning
structures~\cite{HUTH}. An application of our results for more
general cases should take into account that they may be
supplemented by consideration of the complex penetration length
and the quasiparticle contribution in the way as it was made in
the papers~\cite{CLEM, VORTEX}.


\end{document}